# Magnetron Sputtering Formation of Nanoparticles from Natural Olivine Rock for Atmospheric $CO_2$ Capture

Brendan Matulis[1 2], Greyson Wells[1 2], Stefania Crespi[3], Nicola Rotiroti[3], Alessandro Podestà[1], Yuwei Zhang[4], Alberto Calloni[4], Gianlorenzo Bussetti[4], Fernando Cámara[3] and Marcel Di Vece[1 2 5*]

[1] *Physics Department "Aldo Pontremoli", Università degli Studi di Milano, Via Celoria 16, 20133, Milan, Italy*

[2] *Interdisciplinary Centre for Nanostructured Materials and Interfaces (CIMaINa)*

[3] *Dipartimento di Scienze della Terra "A. Desio", Università degli Studi di Milano, Via Luigi Mangiagalli 34, 20133, Milan, Italy*

[4] *Physics Department, Politecnico di Milano, Piazza Leonardo Da Vinci, 20133, Milano, Italy*

[5] *AngelRock BV, Science Park 608, 1098 XH, Amsterdam, The Netherlands*

[*] *Corresponding author: marcel.divece@unimi.it*

# Abstract

The two-birds-one-stone mineralization of $CO_2$ by olivine, is a promising method to both capture carbon directly from the atmosphere and at the same time locking it for storage or utilization. Converting olivine to the nanoscale considerably enhances the kinetics without the need for high temperatures or pressures. Here we present the fabrication of olivine nanoparticles from a natural rock that were fabricated in a gas aggregation magnetron nanoparticle generator. The nanoparticle yield was optimized by enhancing the argon plasma sputter plasma by hydrogen introduction and varying the aggregation distance. The hysteresis of the argon sputter plasma with respect to power is a promising property towards energy efficiency. The formation of well-defined olivine nanoparticles and their subsequent absorption of atmospheric $CO_2$ was confirmed by a suite of techniques. The olivine sputter target surface revealed an intricate interplay between the sputter plasma and olivine composition in terms of crystallinity and morphology. More broadly, this work forms the next step in the practical application of olivine nanoparticles for economical carbon capture and storage, it also is the starting point for the use of this specific nanoparticle technology for mineral-to-nanoparticle conversion.

# 1. Introduction

Anthropogenic climate change is driving research in carbon-neutral energy generation as well as greenhouse gas removal methods such as direct air capture (DAC)[1–3]. Recently, carbon sequestration has become economically incentivised via the carbon trading market[4,5], motivating commercial carbon sequestration initiatives[6,7]. Unfortunately, most of the considered mineralisation technologies still require energy intensive procedures, such as the use of high temperatures and pressures, due to the slow kinetics which depend on temperature, humidity, and pressure[8].

A range of carbon capture and storage methods are being explored such as $CO_2$ amine scrubbing[9], of which mineralisation is probably the oldest large scale and most intensely investigated over recent decades[10–13]. The efficiency, abundance, and availability of $CO_2$-reactive minerals has inspired scientists and technologists to pursue $CO_2$ mineralisation as a means of mitigating anthropogenic climate change[14]. The interaction of $CO_2$ with a wide variety of minerals[4,6], including olivine[12], is a part of Earth's deep carbon cycle, a geological process spanning Earth's history. These interactions include the formation of stable carbonate compounds such as magnesium carbonate (magnesite or $MgCO_3$), thus sequestering atmospheric $CO_2$ in mineral phases over geological time scale, locking atmospheric $CO_2$ into solid mineral phases.

Olivine is widely acknowledged as a promising mineral for the efficient sequestration of $CO_2$[12,15,16] due to its relatively high $CO_2$ absorption capacity (~1:1) and its abundance on earth, formed from outer space olivine[17–19]. However, a major limiting factor is surface passivation: as carbonate products form, they create a coating that inhibits further reaction by blocking access to fresh olivine surfaces[20]. Continuous carbonation therefore relies on weathering processes that disrupt or remove this layer, exposing the unreacted material beneath[15,21].

Most methods, such as high-energy ball milling[22–24] or chemical methods, are often energy-intensive, time-consuming, and ineffective in controlling particle shape[25]. Although recent work has already demonstrated that converting the minerals to the microscale enhances $CO_2$ capture[12,16,26–28], energy intensive methods can be circumvented at smaller nanoscales, as the chemical reactions occur spontaneously and efficiently due to the increased surface to volume ratio[27,29]. Therefore, there is a clear motivation to seek new alternative, scalable, and energy-efficient methods to produce well-defined nanoscale olivine materials for the economically viable capture and storage of atmospheric $CO_2$.

In this work, olivine nanoparticles were generated from a natural olivine rock with a gas aggregation magnetron sputtering nanoparticle generator[30,31] (as recently demonstrated in a proof-of-principle demonstration using a synthetic olivine sputter target[29]). To our knowledge, this represents the first successful application of magnetron sputtering of a natural rock, establishing it as a viable method for the conversion of all types of minerals to nanoparticles. Although improving nanoparticle yield, which currently stands at tens of g/day[32] and reducing energy consumption remain key challenges, considerable improvements have recently been achieved by geometric modifications and careful gas inlet positioning, leading to nanoparticle yield increases of up to sixteen fold[33,34]. Even larger nanoparticle flux improvements were achieved by magnetron magnet configuration engineering, with yields improving up to 160-fold through enhanced dimer formation[34]. Furthermore, magnetron designs such as the cylindrical configuration with rotating magnetic circuitry produces 19 mg/h[35] and a 1.2 yield increase was obtained by the rotating tilted unbalanced yoke magnet[36], while the deposition rate of the Cr coating from the hot solid magnetron target increases 20 times[37]. A recent study on Nylon nanoparticle deposition with a semi-hollow cathode magnetron gas aggregation sputter source yielded deposition rates (up to 1 µm/s)[38], which would translate to close to 300 kg/day for olivine, approaching commercial targets. These novel magnetron sputter designs and process improvements[39], combined together and with upscaling may achieve the required nanoparticle fluxes exceeding kg/h. Considering that for each ton of $CO_2$ roughly a ton of absorbing olivine is required[4], this places it within achievable boundaries.

In addition, the interplay between the natural olivine rock and the argon sputter plasma were studied and optimized in this work. Introducing hydrogen mixed with impurities ($O_2$, $N_2$ etc) from air flushing considerably improved the sputter process. The olivine nanoparticles were characterised with a suite of techniques. Carbonate formation upon exposure to $CO_2$ was confirmed by X-ray photoelectron spectroscopy (XPS). Moreover, this carbonation process was achieved at ambient conditions, which is promising for economically viable carbon capture and storage. This work is therefore a next step in the study of the conversion of olivine to nanoparticle form, further paving the way to the economical sequestration of atmospheric $CO_2$. In principle, the stored carbon could then be used in practical applications[40,41], such as $CO_2$-based fuels[42] and concrete building materials[43]. Beyond $CO_2$ capture, this work establishes magnetron sputtering as a promising process for nanoparticle synthesis from naturally sourced minerals.

# 2. Results and Discussion

## 2.1 Magnetron sputtering nanoparticle yield with air flush and hydrogen

The RF magnetron sputtering of the natural olivine target yielded a very low nanoparticle flux (<1 Å/s) (Figure 3a) when pure argon and a high vacuum (~$10^{-8}$) background pressure were used. Moreover, the magnetron became increasingly unstable with often high reflected power bursts that made operation difficult up from ~60W, indicating an instable plasma. The initial sputtering yielded nanoparticles as shown in Figure 2b, which became impossible to repeat for subsequent runs, likely due to an erosion effect, in which the softer olivine parts became consumed and the harder part remained. The nanoparticle size distribution as shown in Figure 2c reveals an average size of 3.0±1.5nm, which is much smaller than that obtained by sputtering of a synthetic olivine sputter target[29]. The olivine nanoparticles appear faint, likely due to a lack of crystallinity, as observed by electron diffraction and low Z-number of the elements involved[44] with the EDS analysis in Figure 2d confirming the presence of olivine mineral constituents.

Impurities were introduced into the vacuum by flushing the chamber with 30-40 sccm air before magnetron operation. These impurities served as nucleation centres, considerably increasing nanoparticle yield as reported before[45–47]. The relation between the nanoparticle yields (as measured by QCM) and the upward and downward applied magnetron power is shown in Figure 3b-c. Without air flush, a very low flux was measured, with the highest values of only ~0.02 Å/s. In contrast, approximately 10 minutes of air flush improved the overall flux, reaching a peak of 0.07 Å/s at maximum power of 111W (Figure 3a). Flushing for an hour enabled stable flux output and a peak flux of ~0.19 Å/s, representing a 950% increase over no air flush, and a 271% increase over the 10 minute air flush.

By introducing 15% hydrogen to the argon sputter gas, the sputter yield for natural olivine was enhanced from a negligible rate to ~1.2 Å/s (Figure 3b). In contrast to impurity insertion, adding hydrogen to the argon plasma resulted in a large increase in the number of energetic ions hitting the cathode, dramatically increasing the physical sputtering[48,49]. Besides altering the nucleation kinetics of nanoparticles in the aggregation zone, the chemistry between the hydrogen and sputtered atomic species inside the reaction chamber may lead to 1) oxygen scavenging and reduction of iron oxides, partially reducing $Fe^{2+}/Fe^{3+}$[50], 2) conversion of free oxygen atoms into OH and $H_2O$[51], and 3) passivate Si-containing fragments through Si–H bonding[52]. This potentially produces more oxygen-deficient nanoparticles rather than extensive hydride formation[51], and possibly the final olivine stoichiometry as well.

The relation between the nanoparticle yields and the upward and downward applied magnetron power (using 15% $H_2$) as shown in Figure 3b, exhibits a clear hysteresis effect. Increasing the magnetron power to 160 W (~maximum power capacity) increases the olivine nanoparticle flux to ~1.2 Å/s. With a subsequent decrease of the power, a nanoparticle flux was maintained at a stable plateau of ~0.9 Å/s, which then gradually drops to near zero flux as the power decreases from 80 W to close to 65 W. This hysteresis indicates that once the argon sputter plasma has been formed (ignited), its stability allows the power to be reduced by about 50% while also maintaining full operational capacity, leading to considerable energy savings[53]. Because the nanoparticle flux is not zero, the argon sputter plasma must still be present at even lower powers, providing a dramatically increased nanoparticle flux of ~1.2 Å/s starting from ~33W, and an abrupt peak of ~10 Å/s at ~2W. This low power nanoparticle flux was only possible after the argon plasma was created at a higher power. It is possibly caused by the transformation of a more confined plasma at the sputter target to a new, less confined plasma mode at the lowest powers, providing a route to further energy savings. It is known that upon magnetron power switch-off a burst-like release of nanoparticles, previously trapped in the plasma, is observed. However, high nanoparticle flux here observed at low power could be sustained up to a minute, and therefore forms potential for future studies beyond this proof-of-principle work.

### 2.2 Olivine nanoparticle characterisation by TEM and AFM

The olivine nanoparticles were deposited on TEM grids to inspect their size and morphology. For the two aggregation distances (45 mm and 70 mm) and gas flow of Ar with 15% $H_2$, representative TEM images are shown in Figure 4a-b with clear differences. The 45 mm aggregation distance in Figure 4a, shows aggregates of distinct nanoparticles, which resemble the previously reported "cauliflower" structures[54]. However, the nanoparticles deposited with the 70 mm aggregation distance (Figure 4b) were less defined, often with a slight contrast difference around a central darker spot, suggesting easily deformable particles. This was likely caused by the presence of silicon hydroxides, which may have resulted from the mixing of oxygen and hydrogen in the reaction chamber. This difference in morphology may, result from the larger available space between the magnetron and exit aperture that reduces the hydrogen concentration, hence hydroxide formation. The measurement of the nanoparticle sizes in Figure 5c show that the 45 mm and 70 mm aggregation distances results in sizes of 14±4 nm and 4±1 nm, respectively. The slightly smaller nanoparticle size for the 70 mm aggregation distance likely results from a reduced concentration of sputtered species, due to the increased volume which hinders the growth of particles.

The sizes of the olivine nanoparticles that were fabricated with the 15% $H_2$ included in the argon sputter gas were also measured by measuring the height of the nanoparticles deposited on silicon wafers (photograph in Figure 3c) as obtained by AFM (Figure 5 a-b), revealing average sizes of 21±9 nm and 10±4 nm for the aggregation length 45 mm and 70 mm, respectively. The smaller nanoparticle size for the 70 mm aggregation length, indicates a certain unpredictability in the nanoparticle sizes produced with the longer aggregation distance, which should allow the growth of larger nanoparticles —but only when sufficient atomic supply is available. The latter strongly depends on the sputtering yield; therefore, such low concentrations could be reached that further growth of the nanoparticles within the aggregation zone is strongly reduced. The nanoparticle size with the 45 mm aggregation distance and the use of 15% $H_2$ is distinctly larger than the nanoparticles sputtered from the pristine target (Figure 2b-c) due to a more efficient sputtering process providing more atomic supply. The smallest olivine nanoparticle sizes in this work using $H_2$ are only slightly larger compared to laser ablation formed olivine nanoparticles, which are on average 4-9 nm in size[55]. The markedly smaller nanoparticle size as measured by TEM, compared to the measurements by AFM, suggests that the latter measures more aggregated particles, perhaps due to the larger coverage which enables more particles to coalesce on the surface.

EDX measurements (within TEM) on the olivine nanoparticles (Figure 5d) consistently show the presence of all elements from the original olivine mineral, except for the exact stoichiometry, indicating a near-complete conversion from rock to nanoparticle. The substantial presence of magnesium allows for the carbonate formation, and therefore supports its potential technological use in $CO_2$ sequestration. The presence of calcium is likely due to the momentary sputtering of a calcium-rich portion of the sputter target, such as the amphibole tremolite ($Ca_2Mg_5Si_8O_{22}(OH)_2$) grains in equilibrium with the olivine foam microstructure (M2)[56]. This plausible source for Ca is likely not detectable by PWXRD because the amount is < 1-2 % modal.

### 2.3 Olivine nanoparticle carbonation evidence from XPS

An investigation by XPS was conducted on the olivine nanoparticles deposited on silicon wafers (Figure 3c), with the measured elemental composition shown in Table 2. The two samples fabricated with 45 mm and 70 mm aggregation distance have an overall similar composition. The stoichiometry was evaluated on optically uniform layers. However, slight variations were observed when sampling different regions of the same surface (not shown) and between different samples. This is particularly evident for the relative Si intensity, likely due to local exposure of uncovered substrate regions. The Mg to Fe ratio is about 5, indicating a predominance of Mg silicates in the olivine nanoparticles. Germanium is also detected in trace amounts in agreement with EDX analysis, originating from the reactor wall. The oxygen signal can be roughly matched with the oxide stoichiometry expected from the cation concentrations, whereas a large amount of carbon is detected in both samples. This is attributed to the adsorption of adventitious carbon species, mainly airborne hydrocarbons.

For a proper investigation of the surface reactivity with $CO_2$, we refer to the line shape analysis of the photoemission features reported in Figure 6, which shows representative spectra collected from the as-received samples (45 and 70 mm aggregation distances). The spectra are normalised to give approximately the same intensity for the Mg photoemission features. In addition, the binding energy (BE) scale is adjusted by setting the main C 1*s* component, attributed to adventitious carbon, to the reference value of 284.8 eV[57] in order to compensate for differential charging and final-state effects.

Panel (a) in Figure 6 shows the Fe spectral region, characterized by the spin–orbit-split Fe 2$p_{3/2}$ (low-BE) and Fe 2$p_{1/2}$ (high-BE) components. The Fe 2$p_{3/2}$ line-shape displays a main peak at about 710 eV and a satellite feature at 715.5 eV marked by vertical lines. These features are characteristic of Fe(II) species,[58] as expected from natural olivine and Fe-carbonate[55,59]. In panel (b) of Figure 6, a single feature peaking at about 532 eV is observed, assigned to O 1*s* photoemission. This BE value is consistent with either Si-rich phases or carbonates, as further discussed in the literature[60–62]. The C 1*s* line shape in panel (c) is very similar for both samples and consists of a prominent peak due to adventitious carbon species and a side peak at about 289 eV that is usually associated with oxidized carbon species related to $CO_2$ uptake and carbonation reactions[63–65].

The wider scan in panel (d) includes the photoemission signals from Si, Mg, and Fe, with the latter contributing to the minor feature peaking at 54.5 eV (photoemission from Fe 3*p* orbitals). The BE position of the Mg 2*s* and Mg 2*p* features, at 89 eV and 50 eV respectively, is consistent with literature values for Mg silicates[61,66] and is expected to be only weakly affected by changes in the Mg chemical state[61]. The BE difference observed for the Si 2*p* feature is instead likely due to the aforementioned overlap between the photoemission from the olivine film and from the native Si oxide covering the substrate surface.

A quantitative estimate of carbonate species can be obtained by normalising the carbonate carbon concentration to the total concentration of Fe and Mg species. The intensity of the carbonate component, shown as the grey area in Figure 6c, is obtained by numerical fitting of the experimental line shape[64]. In this definition, a value of one indicates complete conversion of olivine into Mg and Fe carbonates. Both samples show generally high carbonation ratios, up to 80%, with some variability also related to the sample history. These results are consistent with the previous work on synthetic olivine[64].

Sample sputtering, performed with an approximate fluence of $10^{16}$ $Ar^+/cm^2$, a kinetic energy 1.5 keV, and grazing incidence, resulted in a measurable decrease in the C 1s intensity, together with changes in the line shape of other photoemission features. Figure 7 focuses on the sample with 45 mm aggregation length and reports the spectral evolution in the O 1*s* [panel (a) of Figure 7] and C 1*s* [panel (b)] BE regions from the as-received condition to the sputtered state, followed by the result of overnight exposure (~14h) to approximately 3 mbar of pure $CO_2$ (*i.e.* about one order of magnitude higher than the $CO_2$ partial pressure in ambient air).

The O 1*s* line develops a shoulder, marked with an asterisk in Figure 7a, at a BE close to the value of 531 eV, reported in the literature for Mg-rich olivine.[66] This observation is consistent with partial removal of the topmost surface layers by sputtering, which partly uncovers the underlying unreacted oxide. Although complete removal of the surface layer is not achieved, and the spectral changes upon $CO_2$ exposure remain limited, further insight into olivine carbonation can be obtained from the inspection of difference spectra (exposed − sputtered). In the O 1*s* region, the spectral changes [panel (a), bottom spectra] fall within the main O photoemission feature. In the C 1*s* region, the carbonate-related signal appears in the 289–290 eV BE range [panel (b), bottom spectra]. As observed in the previous work on synthetic olivine, $CO_2$ interaction in a controlled atmosphere yields a C feature at larger BEs, whereas the lower BE retrieved in the as-received samples is indicative of a slightly altered carbon chemical environment, possibly because of interaction with moisture[64]. An enhancement in the intensity of the side peak (carbonate peak) and the main peak (peak intensity ratios) is observed on transitioning from the “sputtered” to the “exposed” conditions. It is possible to quantify: the carbonate peak is enhanced by about +150%, (computed as $100 * \frac{(I_{exp} - I_{sputt})}{I_{sputt}}$ ), while the enhancement in the main peak is lower (about +40%), consistent with the visual inspection of Figure 7. This points to a ready reaction of the fresh olivine nanoparticle surface after sputtering with the CO2 forming a carbonate. Post-sputtering AFM imaging of the nanoparticles resulted in considerably reduced sizes of 14.4 ±2 nm and 7.5 ±2 nm for the 45 mm and 70 mm aggregation distances, respectively, confirming the ability to expose a pristine olivine surface for $CO_2$ absorption.

### 2.4 Olivine sputter target characterisation

After a combined several hours of total operation, the surface of the olivine sputter target erosion was inspected both visually and by SEM. Form the comparison of the pristine olivine sputter target (Figure 8a) and the eroded sputter target (Figure 8b), reveals that the colour changed from the characteristic green to grey/black, indicating the development of magnetite ($Fe_3O_4$), maghemite (γ-$Fe_2O_3$), or hematite (α-$Fe_2O_3$), all of which appear dark grey to black. Since the back of the olivine sputter target exhibited the same dark colour, it is likely that these changes were induced by the heat produced during magnetron operation. An analysis of the holes, of which the size was estimated by approximating the measured area by a circle, showed sizes of 0.2±0.1mm, which could correspond to the size of softer parts of the olivine.

The plasma etching of the olivine sputter target left distinct geometrically-shaped voids at positions which likely corresponded to crystals with comparatively higher sputter yield due to lower chemical bonding (Figure 10 a-c). In the outer ring, protruding truncated pyramidal crystals were visible (Figure 10d), also found as truncated orthorhombic bipyramids from chemical etching[67]. The Kitt dunite is mylonitized, i.e. it was highly deformed in plastic regime developing areas with crystals, which strongly reduce their size, forming olivine crystallites as truncated orthorhombic bipyramids when it can develop crystal faces. These remaining surface features probably remained due to their lower sputter yield caused by a highly crystalline structure (and hence a stronger chemical bonding than the surrounding olivine). EDS analysis on the truncated pyramidal structures provided atomic concentrations for Mg: 24%, Fe: 1.7%, Si: 13% and O: 55%, which were very similar to the concentrations measured elsewhere on the target surface.

By visual inspection of the eroded olivine sputter target, three differently coloured concentric rings were identified, as indicated in Figure 9. The rings originate from the magnetic field generated by a strong magnetic torus, directing argon primarily towards the centre ring to optimize the sputter process REF. From each of these areas, some fragments were removed and inspected by SEM, revealing that the central area contained redeposited (or back-flowed) nanoparticles[68] (Figure 11a) with a mean particle size of 60±7 nm (Figure 11b), which is at the upper size range as measured by AFM. EDS analysis shows that the redeposited nanoparticles contain mainly iron which, due to its magnetic properties, can be directed by the magnetron's magnetic field, causing them to be redirected back onto the sputter target surface. This is likely a continuous process, and redeposited nanoparticles that were measured likely landed just before power switch-off. Redeposition of plasma-sputtered species is commonly observed to be an obstacle to sputter yield [69,70], and therefore presents an opportunity for efficiency improvement.

EDS measurements (in SEM) on some areas of the sputter target (not shown) result in atomic concentrations for iron up to ~30% and simultaneously for silicon and oxygen down to a few percent, indicating that the post-sputtering surface composition of the olivine has changed due to the redeposition of species and particles. However, because the EDS analysis was not performed on a polished surface, nanoscale roughness could affect the measurement[71].

# 3. Conclusions

The successful conversion of natural olivine rock to nanoparticles by magnetron sputtering represents a promising route for general mineral-to-nanoparticle conversion. The introduction of hydrogen to the argon sputter gas was critical for achieving non-negligible nanoparticle fluxes. Although in this proof-of-concept work, the aggregation distance and air flushing provided parameters to optimize the olivine sputtering process, future studies in which such parameters can be explored in detail, may result in a deeper understanding of the process . The deposited nanoparticles have a high magnesium concentration, which is favourable for $CO_2$ capture. Analysis of the flux versus the applied magnetron power reveals that olivine nanoparticles could be produced at nearly ten times lower power due to a hysteresis effect that dependents on magnetron power —a promising feature for efficient carbon capture and storage. Elemental analysis of the olivine nanoparticles confirmed near-complete conversion of the rock into nanoparticles, i.e. all original elements were retained in the nanoparticles. The presence of substantial amounts of magnesium allowed for carbonate formation at ambient conditions as confirmed by XPS. The small nanoparticle sizes are ideally suited for complete carbonate conversion, and therefore this work, using magnetron sputtering, provides a significant step towards their economically viable use in commercial carbon sequestration.

# 4. Experimental Section

The samples used in this study are natural dunite samples coming from the Kittelfjäll spinel peridotite (KSP), a fragment of the lithospheric mantle preserved as an isolated body within high-grade metamorphic crustal rocks from the Seve Nappe Complex (SNC) in southern Västerbotten, central Sweden (Figure1). KSP is an orogenic peridotite, and it is characterized by a well-developed penetrative compositional layering. This layering is defined by highly depleted dunite, with olivine Mg# (100 × Mg/(Mg + Fe)) ranging from 92.0 to 93.5, and harzburgite with slightly lower Mg# values (91.0–92.5)[56,72]. The sample form this locality corresponds to dunite type M2 (olivine foam microstructure). In these rocks, larger crystals (400 to 2000 µm) are surrounded by smaller crystals (200 to 400 µm). The latter are optically strain-free and characterized by straight grain boundaries that meet in triple point junctions defined by 120° angles between olivine–olivine grain boundaries. Small rock fragments were initially crushed with a mortar and pestle.

Powder X-ray diffraction was performed by a High-resolution X PANalytical X'Pert pro diffractometer with X'Celerator detector. We used $CaF_2$ as internal standard and CuKα as radiation (operating at 40 mA and 40 kV). Data was collected at room–temperature in the 2θ 5.02–90°, with 0.033° step size, collecting for 100.33 sec using a fixed divergence slit (0.25°). Powder XRD analysis of the rock from which the olivine sputter target of 2 inch diameter was obtained, showed it to be almost entirely constituted by olivine with a small amount of brucite ($Mg(OH)_2$). Chromite was also found in these samples with modal percent less than 1% and with a fairly constant composition showing extremely high Cr# (73.1–84.3) and low Mg# (28.7–41.6). The bulk sample composition was obtained by XRF analysis (Table 1). Standardless semiquantitative XRF data collections was performed on a Rigaku Supermini200. Samples were prepared as fused beads with a flux consisting of 66% lithium tetraborate and 34% lithium metaborate, and 0.5% lithium bromide as a release agent.

The nanoparticles were synthesized using a gas aggregation magnetron sputtering nanoparticle generator (NC200U-B, Oxford Applied Research Ltd.) based on the work by Haberland et al.[30,31] in which the natural olivine rock was placed after cutting it into the required 2 inch diameter and 2 mm thickness disk-like shape (Figures 2a and 8a-b). The magnetron was operated in radio frequency (RF) mode at less than 165 W, and using argon as both the sputtering plasma gas and carrier gas within the gas aggregation chamber with a flow rate ~30-55 sccm with operating pressures as measured close to the pump connection of ~$10^{-3}$ mbar. Hydrogen (15%) was added to the argon sputter gas, which resulted in considerable flux increases and provided magnetron stability[48]. Under these conditions, the deposition rates reached up to ~10 Å/s as determined by a quartz crystal microbalance (QCM) set to the $SiO_2$ mass. Although a QCM slightly under estimates the deposition rate of nanoparticles as compared to well-defined atomic deposition due to its reduced contact area, it remains the method of choice[73,74]. Olivine nanoparticle flux as measured by QCM was performed as function of applied RF magnetron power during ~10 min experiments, in which the power was changed in small steps (~15W) after stabilisation of the previous step. Throughout the entire fabrication process of ~5 min for the Si wafer, the samples were ~10 cm from the exit aperture and maintained under vacuum at room temperature. All samples were grown on top of a Si substrate covered with native oxide or cupper grids with a-C membrane for transmission electron microscopy (TEM) investigation. A key parameter in the Haberland nanoparticle generator is the aggregation distance: the distance between the magnetron head and reaction chamber exit aperture. This affects the size of the formed nanoparticles[75,76] and was here used to optimize the nanoparticle yield with the two distances 45 mm and 70 mm.

TEM images were acquired using an FEI Tecnai F20 operating at 200 kV and equipped with an S-Twin lens offering a point resolution of 0.24 nm. Elemental composition was analysed by energy-dispersive X-ray spectroscopy (EDS) using an Oxford Instruments Xplore system with an ultrathin window, confirming the presence of all sputter target constituents in the nanoparticles. Nanoparticle size estimations were performed by identifying visible and separate particles. For nanoparticles which were connected, a regions of interest (ROI) was drawn around them by hand in ImageJ, and subsequently the mean diameter was used as a representative of particle size, sometimes by taking the straight lines across their diameters.

The surface morphology of the olivine nanoparticle assembled films were determined by atomic force microscopy (AFM) using a DriveAFM microscope by Nanosurf operated in piezo-driven Tapping Mode in air, equipped with standard rigid probes with a tip radius below 10 nm. AFM raw images were flattened by subtracting up to 1$^{st}$ order polynomials line-by-line to remove scanner and/or cantilever drifts as well as sample tilt. The particle height was measured by drawing line profiles across distinct nanoparticles and measuring the crest-to-trough height

The X-ray photoelectron spectroscopy (XPS) characterization was performed in a custom-built Ultra-High Vacuum (UHV) system (base pressure in the $10^{-11}$ Torr range) equipped with a 150 mm hemispherical analyser from SPECS GmbH and a Mg Kα ($h\nu = 1253.6$ eV) photon source. XPS scans were acquired at a fixed pass energy of 20 eV, yielding an overall Full Width at Half Maximum (FWHM) energy resolution of about 1 eV. The photoelectron spectrometer was equipped with a differentially pumped $Ar^+$ sputter gun for the *in situ* removal of the topmost surface layers and a load-lock chamber at high vacuum conditions (base pressure in the $10^{-8}$ Torr range) that could be flooded with pure $CO_2$ gas through a leak valve[77].

Elemental concentrations were retrieved from experimental XPS intensities after normalisation to element- and transition-specific photoelectric cross sections[78] and to approximate parametrizations accounting for the finite analysis depth and transmission of the photoelectron analyzer[79,80]. The information depth is roughly evaluated as three times the inelastic mean free path ($\lambda_m$) of the detected photoelectrons in the sample. Assuming $\lambda_m \approx 3$ nm,[80], this gives an information depth of about 9 nm. An accuracy of about 10% of the reported values should be considered, accounting for the approximations used.

A morphological examination of the olivine sputter target at different positions after scraping was performed using a JEOL JSM-IT500 La scanning electron microscope (SEM) to investigate the result of the deposition. Since the scrapes sample had an insulating surface, the sample itself was sputter deposited with a layer of few nm of gold. The images were acquired using 20 kV to accelerate the primary electrons.

### Author contributions

MDV conceived and supervised the work, made the samples and wrote the manuscript. BM and GW assisted with experiments and performed data analysis. FC and NR contributed to sample preparation and TEM and EDS analysis. AP performed AFM and analysed the data. GB, AC and YZ performed XPS and analysed the data. SC performed SEM and EDS on the sputter target. All authors contributed to the manuscript by discussion and editing.

## Acknowledgement

The authors thank M. Giglio (University of Pavia) who provided the sample from KPS and Chiara Mangano (University of Milan) for XRD and XRF analyses. M. Perfido (University of Milan) is thanked for drilling and the cutting of the dunite targets. Megan Farrington is thanked for her assistance during depositions. MDV appreciates the stimulating discussions with Adam Obrusník and Kristína Šťastná.

# References

(1) Can Carbon Removal Become a Trillion-Dollar Business? *The Economist* **2023**, 5.

(2) Sanz-Pérez, E. S.; Murdock, C. R.; Didas, S. A.; Jones, C. W. Direct Capture of CO2 from Ambient Air. *Chem. Rev.* **2016**, *116* (19), 11840–11876. https://doi.org/10.1021/acs.chemrev.6b00173.

(3) Bui, M.; Adjiman, C. S.; Bardow, A.; Anthony, E. J.; Boston, A.; Brown, S.; Fennell, P. S.; Fuss, S.; Galindo, A.; Hackett, L. A.; Hallett, J. P.; Herzog, H. J.; Jackson, G.; Kemper, J.; Krevor, S.; Maitland, G. C.; Matuszewski, M.; Metcalfe, I. S.; Petit, C.; Puxty, G.; Reimer, J.; Reiner, D. M.; Rubin, E. S.; Scott, S. A.; Shah, N.; Smit, B.; Trusler, J. P. M.; Webley, P.; Wilcox, J.; Mac Dowell, N. Carbon Capture and Storage (CCS): The Way Forward. *Energy Environ. Sci.* **2018**, *11* (5), 1062–1176. https://doi.org/10.1039/C7EE02342A.

(4) Sanna, A.; Uibu, M.; Caramanna, G.; Kuusik, R.; Maroto-Valer, M. M. A Review of Mineral Carbonation Technologies to Sequester CO2. *Chem. Soc. Rev.* **2014**, *43* (23), 8049–8080. https://doi.org/10.1039/C4CS00035H.

(5) Olajire, A. A. A Review of Mineral Carbonation Technology in Sequestration of CO2. *J. Pet. Sci. Eng.* **2013**, *109*, 364–392. https://doi.org/10.1016/j.petrol.2013.03.013.

(6) Snæbjörnsdóttir, S. Ó.; Sigfússon, B.; Marieni, C.; Goldberg, D.; Gislason, S. R.; Oelkers, E. H. Carbon Dioxide Storage through Mineral Carbonation. *Nat. Rev. Earth Environ.* **2020**, *1* (2), 90–102. https://doi.org/10.1038/s43017-019-0011-8.

(7) Michaelowa, A.; Honegger, M.; Poralla, M.; Winkler, M.; Dalfiume, S.; Nayak, A. International Carbon Markets for Carbon Dioxide Removal. *PLOS Clim.* **2023**, *2*, e0000118. https://doi.org/10.1371/journal.pclm.0000118.

(8) Kelemen, P.; Aines, R.; Bennett, E.; Benson, S.; Carter, E.; Coggon, J.; de Obeso, J.; Evans, O.; Gadikota, G.; Dipple, G.; Godard, M.; Harris, M.; Higgins, J.; Johnson, K.; Kourim, F.; Lafay, R.; Lambart, S.; Manning, C.; Matter, J.; Michibayashi, K.; Morishita, T.; Noël, J.; Okazaki, K.; Renforth, P.; Robinson, B.; Savage, H.; Skarbek, R.; Spiegelman, M.; Takazawa, E.; Teagle, D.; Urai, J.; Wilcox, J.; Oman Drilling Project Phase 1 Sci. In Situ Carbon Mineralization in Ultramafic Rocks: Natural Processes and Possible Engineered Methods; Marieni, C., Voigt, M., Snaebjornsdottir, S., Oelkers, E., Eds.; 2018; Vol. 146, pp 92–102. https://doi.org/10.1016/j.egypro.2018.07.013.

(9) Rochelle, G. T. Amine Scrubbing for CO2 Capture. *Science* **2009**, *325* (5948), 1652–1654. https://doi.org/10.1126/science.1176731.

(10) Mora Mendoza, E. Y.; Sarmiento Santos, A.; Vera López, E.; Drozd, V.; Durygin, A.; Chen, J.; Saxena, S. K. Iron Oxides as Efficient Sorbents for CO2 Capture. *J. Mater. Res. Technol.* **2019**, *8* (3), 2944–2956. https://doi.org/10.1016/j.jmrt.2019.05.002.

(11) Duan, Y.; Sorescu, D.; Luebke, D. Efficient Theoretical Screening of Solid Sorbents for CO2 Capture Applications. *Int. J. Clean Coal Energy* **2012**, *01*. https://doi.org/10.4236/ijcce.2012.11001.

(12) Kelemen, P. B.; Matter, J. In Situ Carbonation of Peridotite for CO2 Storage. *Proc. Natl. Acad. Sci.* **2008**, *105* (45), 17295–17300. https://doi.org/10.1073/pnas.0805794105.

(13) Arias, B.; Alonso, M.; Abanades, C. CO2 Capture by Calcium Looping at Relevant Conditions for Cement Plants: Experimental Testing in a 30 kWth Pilot Plant. *Ind. Eng. Chem. Res.* **2017**, *56* (10), 2634–2640. https://doi.org/10.1021/acs.iecr.6b04617.

(14) Keith, D.; Holmes, G.; Angelo, D.; Heidel, K. A Process for Capturing CO2 from the Atmosphere. *Joule* **2018**, *2*. https://doi.org/10.1016/j.joule.2018.05.006.

(15) Kelemen, P.; Benson, S. M.; Pilorgé, H.; Psarras, P.; Wilcox, J. An Overview of the Status and Challenges of CO2 Storage in Minerals and Geological Formations. *Front. Clim.* **2019**, *1*. https://doi.org/10.3389/fclim.2019.00009.

(16) Wang, F.; Dreisinger, D.; Jarvis, M.; Hitchins, T. Kinetics and Mechanism of Mineral Carbonation of Olivine for CO2 Sequestration. *Miner. Eng.* **2019**, *131*, 185–197. https://doi.org/10.1016/j.mineng.2018.11.024.

(17) Ebel, D. S.; Weisberg, M. K.; Beckett, J. R. Thermochemical Stability of Low-Iron, Manganese-Enriched Olivine in Astrophysical Environments. *Meteorit. Planet. Sci.* **2012**, *47* (4), 585–593. https://doi.org/10.1111/j.1945-5100.2012.01347.x.

(18) Grün, E.; Krüger, H.; Srama, R. The Dawn of Dust Astronomy. *Space Sci. Rev.* **2019**, *215* (7), 46. https://doi.org/10.1007/s11214-019-0610-1.

(19) Sanchez, J. A.; Reddy, V.; Kelley, M. S.; Cloutis, E. A.; Bottke, W. F.; Nesvorný, D.; Lucas, M. P.; Hardersen, P. S.; Gaffey, M. J.; Abell, P. A.; Corre, L. L. Olivine-Dominated Asteroids: Mineralogy and Origin. *Icarus* **2014**, *228*, 288–300. https://doi.org/10.1016/j.icarus.2013.10.006.

(20) Béarat, H.; McKelvy, M. J.; Chizmeshya, A. V. G.; Gormley, D.; Nunez, R.; Carpenter, R. W.; Squires, K.; Wolf, G. H. Carbon Sequestration via Aqueous Olivine Mineral Carbonation:  Role of Passivating Layer Formation. *Environ. Sci. Technol.* **2006**, *40* (15), 4802–4808. https://doi.org/10.1021/es0523340.

(21) Pokrovsky, O.; Schott, J. Kinetics and Mechanism of Forsterite Dissolution at 25°C and pH from 1 to 12. *Geochim. Cosmochim. ACTA* **2000**, *64* (19), 3313–3325. https://doi.org/10.1016/S0016-7037(00)00434-8.

(22) Nguyen, P. Q. H.; Zhang, D.; Rapp, R.; Bradley, J. P.; Dera, P. Room Temperature Facile Synthesis of Olivine-Co2SiO4 Nanoparticles Utilizing a Mechanochemical Method. *RSC Adv.* **2021**, *11* (34), 20687–20690. https://doi.org/10.1039/D1RA02760C.

(23) Farina, V.; Gamba, N. S.; Gennari, F.; Garroni, S.; Torre, F.; Taras, A.; Enzo, S.; Mulas, G. CO2 Hydrogenation Induced by Mechanochemical Activation of Olivine With Water Under CO2 Atmosphere. *Front. Energy Res.* **2019**, *7*. https://doi.org/10.3389/fenrg.2019.00107.

(24) Ratnawulan Ratnawulan; Ahmad Fauzi. Synthesis and Characterization of Forsterite (Mg2SiO4) Nanomaterials of Dunite from Sumatera. In *Recent Advances in Pyrolysis*; Hassan Al- Haj Ibrahim, Ed.; IntechOpen: Rijeka, 2019; p Ch. 6. https://doi.org/10.5772/intechopen.85082.

(25) Saunders, R. W.; Plane, J. M. C. A Photo-Chemical Method for the Production of Olivine Nanoparticles as Cosmic Dust Analogues. *Icarus* **2011**, *212* (1), 373–382. https://doi.org/10.1016/j.icarus.2010.12.019.

(26) Turri, L.; Gérardin, K.; Muhr, H.; Lapicque, F.; Saravia, A.; Szenknect, S.; Mesbah, A.; Mastretta, R.; Dacheux, N.; Meyer, D.; Cloteaux, A.; Gérard, A.; Bertucci, S. CO2 Sequestration by Carbonation of Olivine: A New Process for Optimal Separation of the Solids Produced. *Green Process. Synth.* **2019**, *8* (1), 480–487. https://doi.org/doi:10.1515/gps-2019-0016.

(27) Rahmani, O.; Highfield, J.; Junin, R.; Tyrer, M.; Pour, A. B. Experimental Investigation and Simplistic Geochemical Modeling of CO2 Mineral Carbonation Using the Mount Tawai Peridotite. *Molecules* **2016**, *21* (3). https://doi.org/10.3390/molecules21030353.

(28) Dufourny, A.; Julcour, C.; Esvan, J.; Cassayre, L.; Laniesse, P.; Bourgeois, F. Observation of the Depassivation Effect of Attrition on Magnesium Silicates' Direct Aqueous Carbonation Products. *Front. Clim.* **2022**, *4*. https://doi.org/10.3389/fclim.2022.946735.

(29) Iozzia, M. L.; Goto, F.; Podestà, A.; Vecchi, R.; Calloni, A.; Lenardi, C.; Bussetti, G.; Di Vece, M. Olivine Nanoparticles for Fast Atmospheric CO2 Capture at Ambient Conditions. *Part. Part. Syst. Charact.* **2024**, *n/a* (n/a), 2400063. https://doi.org/10.1002/ppsc.202400063.

(30) Haberland, H.; Mall, M.; Moseler, M.; Qiang, Y.; Reiners, T.; Thurner, Y. Filling of Micron-sized Contact Holes with Copper by Energetic Cluster Impact. *J. Vac. Sci. Technol. Vac. Surf. Films* **1994**, *12* (5), 2925–2930. https://doi.org/10.1116/1.578967.

(31) Haberland, H.; Karrais, M.; Mall, M.; Thurner, Y. Thin Films from Energetic Cluster Impact: A Feasibility Study. *J. Vac. Sci. Technol. Vac. Surf. Films* **1992**, *10* (5), 3266–3271. https://doi.org/10.1116/1.577853.

(32) Martínez, L.; Lauwaet, K.; Santoro, G.; Sobrado, J. M.; Peláez, R. J.; Herrero, V. J.; Tanarro, I.; Ellis, G. J.; Cernicharo, J.; Joblin, C.; Huttel, Y.; Martín-Gago, J. A. Precisely Controlled Fabrication, Manipulation and in-Situ Analysis of Cu Based Nanoparticles. *Sci. Rep.* **2018**, *8* (1), 7250. https://doi.org/10.1038/s41598-018-25472-y.

(33) Sanzone, G.; Yin, J.; Cooke, K.; Sun, H.; Lievens, P. Impact of the Gas Dynamics on the Cluster Flux in a Magnetron Cluster-Source: Influence of the Chamber Shape and Gas-Inlet Position. *Rev. Sci. Instrum.* **2021**, *92* (3), 033901. https://doi.org/10.1063/5.0028854.

(34) Coroa, J.; Sanzone, G.; Höltzl, T.; Sun, H.; Janssens, E.; Yin, J. The Influence of Magnetic Field on the Cluster Growth in a Magnetron Sputtering Gas Aggregation Source. *Surf. Coat. Technol.* **2025**, *500*, 131892. https://doi.org/10.1016/j.surfcoat.2025.131892.

(35) Nikitin, D.; Hanuš, J.; Pleskunov, P.; Krtouš, Z.; Ali-Ogly, S.; Tafiichuk, R.; Biliak, K.; Protsak, M.; Valter, J.; Vyskočil, J.; Choukourov, A.; Biederman, H. Novel Gas Aggregation Cluster Source Based on Post Magnetron. *Plasma Process. Polym.* **2021**, *18* (9), 2100068. https://doi.org/10.1002/ppap.202100068.

(36) Iseki, T. Target Utilization of Planar Magnetron Sputtering Using a Rotating Tilted Unbalanced Yoke Magnet. *Vacuum* **2009**, *84* (2), 339–347. https://doi.org/10.1016/j.vacuum.2009.07.010.

(37) Bleykher, G. A.; Borduleva, A. O.; Krivobokov, V. P.; Sidelev, D. V. Evaporation Factor in Productivity Increase of Hot Target Magnetron Sputtering Systems. *Vacuum* **2016**, *132*, 62–69. https://doi.org/10.1016/j.vacuum.2016.07.030.

(38) Solař, P.; Melnichuk, I.; Artemenko, A.; Polonskyi, O.; Kylián, O.; Choukourov, A.; Slavínská, D.; Biederman, H. Nylon-Sputtered Plasma Polymer Particles Produced by a Semi-Hollow Cathode Gas Aggregation Source. *Vacuum* **2015**, *111*, 124–130. https://doi.org/10.1016/j.vacuum.2014.09.023.

(39) Gudmundsson, J. T. Physics and Technology of Magnetron Sputtering Discharges. *Plasma Sources Sci. Technol.* **2020**, *29* (11), 113001. https://doi.org/10.1088/1361-6595/abb7bd.

(40) Kim, C.; Yoo, C.-J.; Oh, H.-S.; Min, B. K.; Lee, U. Review of Carbon Dioxide Utilization Technologies and Their Potential for Industrial Application. *J. CO2 Util.* **2022**, *65*, 102239. https://doi.org/10.1016/j.jcou.2022.102239.

(41) Li, B.; Duan, Y.; Luebke, D.; Morreale, B. Advances in CO2 Capture Technology: A Patent Review. *Spec. Issue Adv. Sustain. Biofuel Prod. Use - XIX Int. Symp. Alcohol Fuels - ISAF* **2013**, *102*, 1439–1447. https://doi.org/10.1016/j.apenergy.2012.09.009.

(42) Kondratenko, E.; Mul, G.; Baltrusaitis, J.; Larrazábal, G.; Pérez-Ramírez, J. Status and Perspectives of CO2 Conversion into Fuels and Chemicals by Catalytic, Photocatalytic and Electrocatalytic Processes. *ENERGY Environ. Sci.* **2013**, *6* (11), 3112–3135. https://doi.org/10.1039/c3ee41272e.

(43) Hepburn, C.; Adlen, E.; Beddington, J.; Carter, E. A.; Fuss, S.; Mac Dowell, N.; Minx, J. C.; Smith, P.; Williams, C. K. The Technological and Economic Prospects for CO2 Utilization and Removal. *Nature* **2019**, *575* (7781), 87–97. https://doi.org/10.1038/s41586-019-1681-6.

(44) Pozsgai, I. Thickness Determination by Measuring Electron Transmission in the TEM at 200kV. *Ultramicroscopy* **1997**, *68* (1), 69–75. https://doi.org/10.1016/S0304-3991(97)00024-7.

(45) Polonskyi, O.; Kylián, O.; Drábik, M.; Kousal, J.; Solař, P.; Artemenko, A.; Čechvala, J.; Choukourov, A.; Slavínská, D.; Biederman, H. Deposition of Al Nanoparticles and Their Nanocomposites Using a Gas Aggregation Cluster Source. *J. Mater. Sci.* **2014**, *49* (9), 3352–3360. https://doi.org/10.1007/s10853-014-8042-5.

(46) Marek, A.; Valter, J.; Kadlec, S.; Vyskočil, J. Gas Aggregation Nanocluster Source — Reactive Sputter Deposition of Copper and Titanium Nanoclusters. *PSE 2010 Spec. Issue* **2011**, *205*, S573–S576. https://doi.org/10.1016/j.surfcoat.2010.12.027.

(47) Owen, C. A.; Podestà, A.; Lenardi, C.; Kadkhodazadeh, S.; Di Vece, M. Photo-Stimulated Hydrogen Desorption from Magnesium Nanoparticles. *Int. J. Hydrog. Energy* **2022**, *47* (81), 34594–34604. https://doi.org/10.1016/j.ijhydene.2022.08.044.

(48) Budtz-Jørgensen, C. V.; Kringhøj, P.; Bøttiger, J. The Critical Role of Hydrogen for Physical Sputtering with Ar–H2 Glow Discharges. *Surf. Coat. Technol.* **1999**, *116–119*, 938–943. https://doi.org/10.1016/S0257-8972(99)00126-7.

(49) Acsente, T.; Stoica, S. D.; Craciun, C.; Mitu, B.; Dinescu, G. Enhancement of W Nanoparticles Synthesis by Injecting H2 in a Magnetron Sputtering Gas Aggregation Cluster Source Operated in Ar. *Plasma Chem. Plasma Process.* **2024**, *44* (6), 2233–2246. https://doi.org/10.1007/s11090-024-10499-z.

(50) Singh, B. B.; Kumar, S.; Krall, E. A.; Goldman, M.; Heo, T. W.; Kortshagen, U.; Bruggeman, P. J. Mechanism of H2 Plasma-Enabled Reduction of Hematite Thin Films. *Int. J. Hydrog. Energy* **2026**, *244*, 155624. https://doi.org/10.1016/j.ijhydene.2026.155624.

(51) Jiménez-Redondo, M.; Carrasco, E.; Herrero, V. J.; Tanarro, I. Chemistry in Glow Discharges of H2/O2 Mixtures: Diagnostics and Modelling. *Plasma Sources Sci. Technol.* **2015**, *24* (1), 015029. https://doi.org/10.1088/0963-0252/24/1/015029.

(52) Seo, J.-M.; Jeong, M.-C.; Myoung, J.-M. Effects of Hydrogen on Poly- and Nano-Crystallization of a-Si:H Prepared by RF Magnetron Sputtering. *J. Cryst. Growth* **2006**, *295* (2), 119–123. https://doi.org/10.1016/j.jcrysgro.2006.07.028.

(53) Berg, S.; Nyberg, T. Fundamental Understanding and Modeling of Reactive Sputtering Processes. *Thin Solid Films* **2005**, *476* (2), 215–230. https://doi.org/10.1016/j.tsf.2004.10.051.

(54) Tang, W.; Eilers, J. J.; van Huis, M. A.; Wang, D.; Schropp, R. E. I.; Di Vece, M. Formation and Photoluminescence of “Cauliflower” Silicon Nanoparticles. *J. Phys. Chem. C* **2015**, *119* (20), 11042–11047. https://doi.org/10.1021/jp511660h.

(55) Murri, M.; Capitani, G.; Fasoli, M.; Monguzzi, A.; Calloni, A.; Bussetti, G.; Malaspina, N.; Campione, M. Laboratory Simulation of Space Weathering on Silicate Surfaces in the Water Environment. *ACS Earth Space Chem.* **2022**, *6* (1), 197–206. https://doi.org/10.1021/acsearthspacechem.1c00349.

(56) Clos, F.; Gilio, M.; van Roermund, H. L. M. Fragments of Deeper Parts of the Hanging Wall Mantle Preserved as Orogenic Peridotites in the Central Belt of the Seve Nappe Complex, Sweden. *Lithos* **2014**, *192–195*, 8–20. https://doi.org/10.1016/j.lithos.2014.01.004.

(57) Morgan, D. J. The Utility of Adventitious Carbon for Charge Correction: A Perspective From a Second Multiuser Facility. *Surf. Interface Anal.* **2025**, *57* (1), 28–35. https://doi.org/10.1002/sia.7360.

(58) Gota, S.; Guiot, E.; Henriot, M.; Gautier-Soyer, M. Atomic-Oxygen-Assisted MBE Growth of a-Fe2O3 on a-Al2O3„0001…: Metastable FeO„111…-like Phase at Subnanometer Thicknesses.

(59) Biesinger, M. C.; Payne, B. P.; Grosvenor, A. P.; Lau, L. W. M.; Gerson, A. R.; Smart, R. St. C. Resolving Surface Chemical States in XPS Analysis of First Row Transition Metals, Oxides and Hydroxides: Cr, Mn, Fe, Co and Ni. *Appl. Surf. Sci.* **2011**, *257* (7), 2717–2730. https://doi.org/10.1016/j.apsusc.2010.10.051.

(60) Wagner, C. D. *NIST X-Ray Photoelectron Spectroscopy (XPS) Database*, 0 ed.; NBS TN 1289; National Bureau of Standards: Gaithersburg, MD, 1990; p NBS TN 1289. https://doi.org/10.6028/NIST.TN.1289.

(61) Schulze, R. K.; Hill, M. A.; Field, R. D.; Papin, P. A.; Hanrahan, R. J.; Byler, D. D. Characterization of Carbonated Serpentine Using XPS and TEM. *Energy Convers. Manag.* **2004**, *45* (20), 3169–3179. https://doi.org/10.1016/j.enconman.2004.02.003.

(62) Luong, N. T.; Veyret, N.; Boily, J.-F. $CO_2$ Mineralization by MgO Nanocubes in Nanometric Water Films. *ACS Appl. Mater. Interfaces* **2023**, *15* (38), 45055–45063. https://doi.org/10.1021/acsami.3c10590.

(63) Dufourny, A.; Julcour, C.; Esvan, J.; Cassayre, L.; Laniesse, P.; Bourgeois, F. Observation of the Depassivation Effect of Attrition on Magnesium Silicates' Direct Aqueous Carbonation Products. *Front. Clim.* **2022**, *4*, 946735. https://doi.org/10.3389/fclim.2022.946735.

(64) Iozzia, M. L.; Goto, F.; Podestà, A.; Vecchi, R.; Calloni, A.; Lenardi, C.; Bussetti, G.; Di Vece, M. Olivine Nanoparticles for Fast Atmospheric $CO_2$ Capture at Ambient Conditions. *Part. Part. Syst. Charact.* **2025**, *42* (1), 2400063. https://doi.org/10.1002/ppsc.202400063.

(65) Shchukarev, A.; Korolkov, D. XPS Study of Group IA Carbonates. *Open Chem.* **2004**, *2* (2), 347–362. https://doi.org/10.2478/BF02475578.

(66) Zakaznova-Herzog, V. P.; Nesbitt, H. W.; Bancroft, G. M.; Tse, J. S.; Gao, X.; Skinner, W. High-Resolution Valence-Band XPS Spectra of the Nonconductors Quartz and Olivine. *Phys. Rev. B* **2005**, *72* (20), 205113. https://doi.org/10.1103/PhysRevB.72.205113.

(67) Wegner, M. W.; Christie, J. M. Preferential Chemical Etching of Terrestrial And Lunar Olivines. *Contrib. Mineral. Petrol.* **1974**, *43* (3), 195–212. https://doi.org/10.1007/BF01134836.

(68) Shelemin, A.; Pleskunov, P.; Kousal, J.; Drewes, J.; Hanuš, J.; Ali-Ogly, S.; Nikitin, D.; Solař, P.; Kratochvíl, J.; Vaidulych, M.; Schwartzkopf, M.; Kylián, O.; Polonskyi, O.; Strunskus, T.; Faupel, F.; Roth, S. V.; Biederman, H.; Choukourov, A. Nucleation and Growth of Magnetron-Sputtered Ag Nanoparticles as Witnessed by Time-Resolved Small Angle X-Ray Scattering. *Part. Part. Syst. Charact.* **2020**, *37* (2), 1900436. https://doi.org/10.1002/ppsc.201900436.

(69) Rossnagel, S. M. Deposition and Redeposition in Magnetrons. *J. Vac. Sci. Technol. A* **1988**, *6* (6), 3049–3054. https://doi.org/10.1116/1.575473.

(70) Strijckmans, K.; Depla, D. Modeling Target Erosion during Reactive Sputtering. *Appl. Surf. Sci.* **2015**, *331*, 185–192. https://doi.org/10.1016/j.apsusc.2015.01.058.

(71) Newbury, D. E. Quantitative Electron Probe Microanalysis of Rough Targets: Testing the Peak-to-Local Background Method. *Scanning* **2004**, *26* (3), 103–114. https://doi.org/10.1002/sca.4950260302.

(72) van Roermund, H. L. M. Eclogites of the Seve Nappe, Central Scandinavian Caledonides. In *The Caledonide Orogen - Scandinavia and related areas*; John Wiley & Sons: Chichester, 1985; Vol. 2, pp 873-886.

(73) Drewes, J.; Ali-Ogly, S.; Strunskus, T.; Polonskyi, O.; Biederman, H.; Faupel, F.; Vahl, A. Impact of Argon Flow and Pressure on the Trapping Behavior of Nanoparticles inside a Gas Aggregation Source. *Plasma Process. Polym.* **2022**, *19* (1), 2100125. https://doi.org/10.1002/ppap.202100125.

(74) Gauter, S.; Haase, F.; Solař, P.; Kylián, O.; Kúš, P.; Choukourov, A.; Biederman, H.; Kersten, H. Calorimetric Investigations in a Gas Aggregation Source. *J. Appl. Phys.* **2018**, *124* (7), 073301. https://doi.org/10.1063/1.5037413.

(75) Khojasteh, M.; Kresin, V. V. Influence of Source Parameters on the Growth of Metal Nanoparticles by Sputter-Gas-Aggregation. *Appl. Nanosci.* **2017**, *7* (8), 875–883. https://doi.org/10.1007/s13204-017-0627-2.

(76) Nikitin, D.; Hanuš, J.; Ali-Ogly, S.; Polonskyi, O.; Drewes, J.; Faupel, F.; Biederman, H.; Choukourov, A. The Evolution of Ag Nanoparticles inside a Gas Aggregation Cluster Source. *Plasma Process. Polym.* **2019**, *16* (10), 1900079. https://doi.org/10.1002/ppap.201900079.

(77) Berti, G.; Calloni, A.; Brambilla, A.; Bussetti, G.; Duò, L.; Ciccacci, F. Direct Observation of Spin-Resolved Full and Empty Electron States in Ferromagnetic Surfaces. *Rev. Sci. Instrum.* **2014**, *85* (7), 073901. https://doi.org/10.1063/1.4885447.

(78) Scofield, J. H. Hartree-Slater Subshell Photoionization Cross-Sections at 1254 and 1487 eV. *J. Electron Spectrosc. Relat. Phenom.* **1976**, *8* (2), 129–137. https://doi.org/10.1016/0368-2048(76)80015-1.

(79) Fairley, N.; Fernandez, V.; Richard-Plouet, M.; Guillot-Deudon, C.; Walton, J.; Smith, E.; Flahaut, D.; Greiner, M.; Biesinger, M.; Tougaard, S.; Morgan, D.; Baltrusaitis, J. Systematic and Collaborative Approach to Problem Solving Using X-Ray Photoelectron Spectroscopy. *Appl. Surf. Sci. Adv.* **2021**, *5*, 100112. https://doi.org/10.1016/j.apsadv.2021.100112.

(80) Jablonski, A. Universal Energy Dependence of the Inelastic Mean Free Path. *Surf. Interface Anal.* **1993**, *20* (4), 317–321. https://doi.org/10.1002/sia.740200409.

## Figures

FIGURE 1

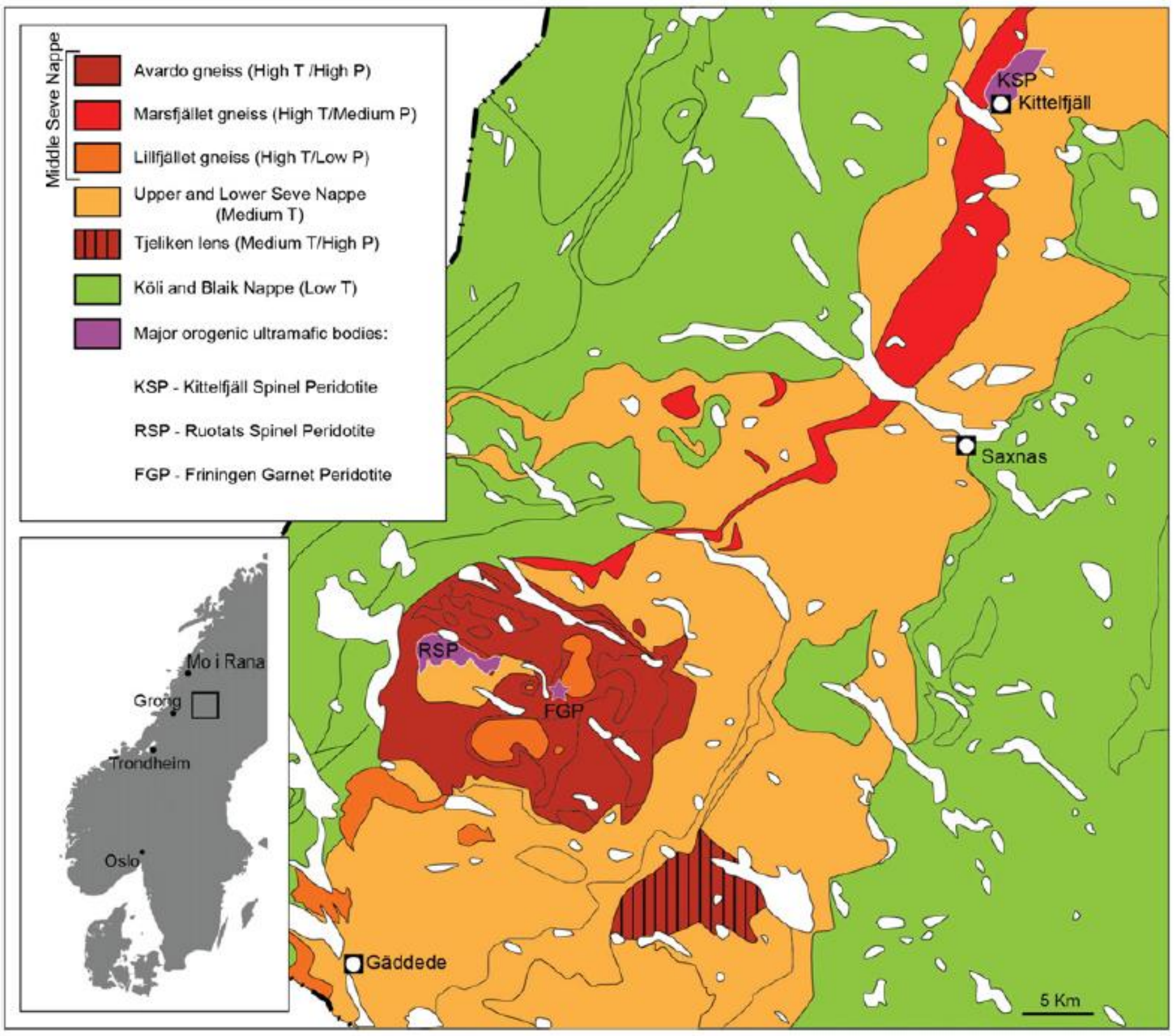

Metamorphic map of the Seve–Köli Nappe Complex in northern Jämtland/Southern-Västerbotten, Central Sweden. The black solid lines present in the background of the map refer to the main geological units outlined on the geological map presented by Van Roermund (1985), reference 72. All rock units on the map generally dip(moderately) to the west/northwest. Copyright Clos et al. (2014) reference 56

TABLE 1

| | Kittelfjäll |
|---|---|
| $MgO$ | 52.43 |
| $Al_2O_3$ | 0.47 |
| $SiO_2$ | 37.07 |
| $K_2O$ | 0.28 |
| $CaO$ | 0.05 |
| $Cr_2O_3$ | 0.56 |
| $MnO$ | 0.12 |
| $Fe_2O_3$ | 8.55 |
| $NiO$ | 0.43 |
| $ZrO_2$ | 0.03 |
| | 100.00% |

The bulk sample composition (%) was obtained by XRF analysis, confirming olivine constituents

FIGURE 2

A)

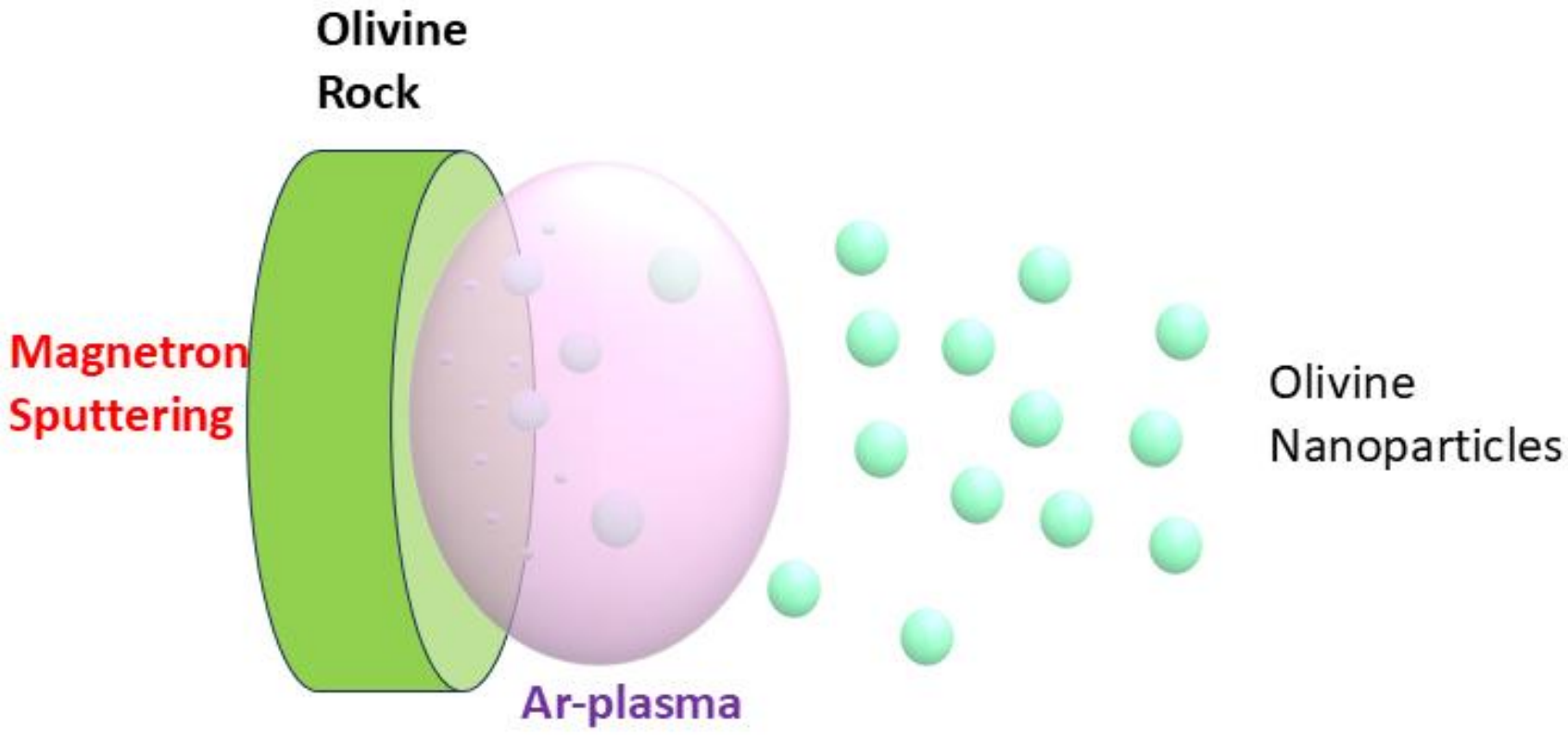
Olivine
Rock
Magnetron
Sputtering
Ar-plasma
Olivine
Nanoparticles

B)

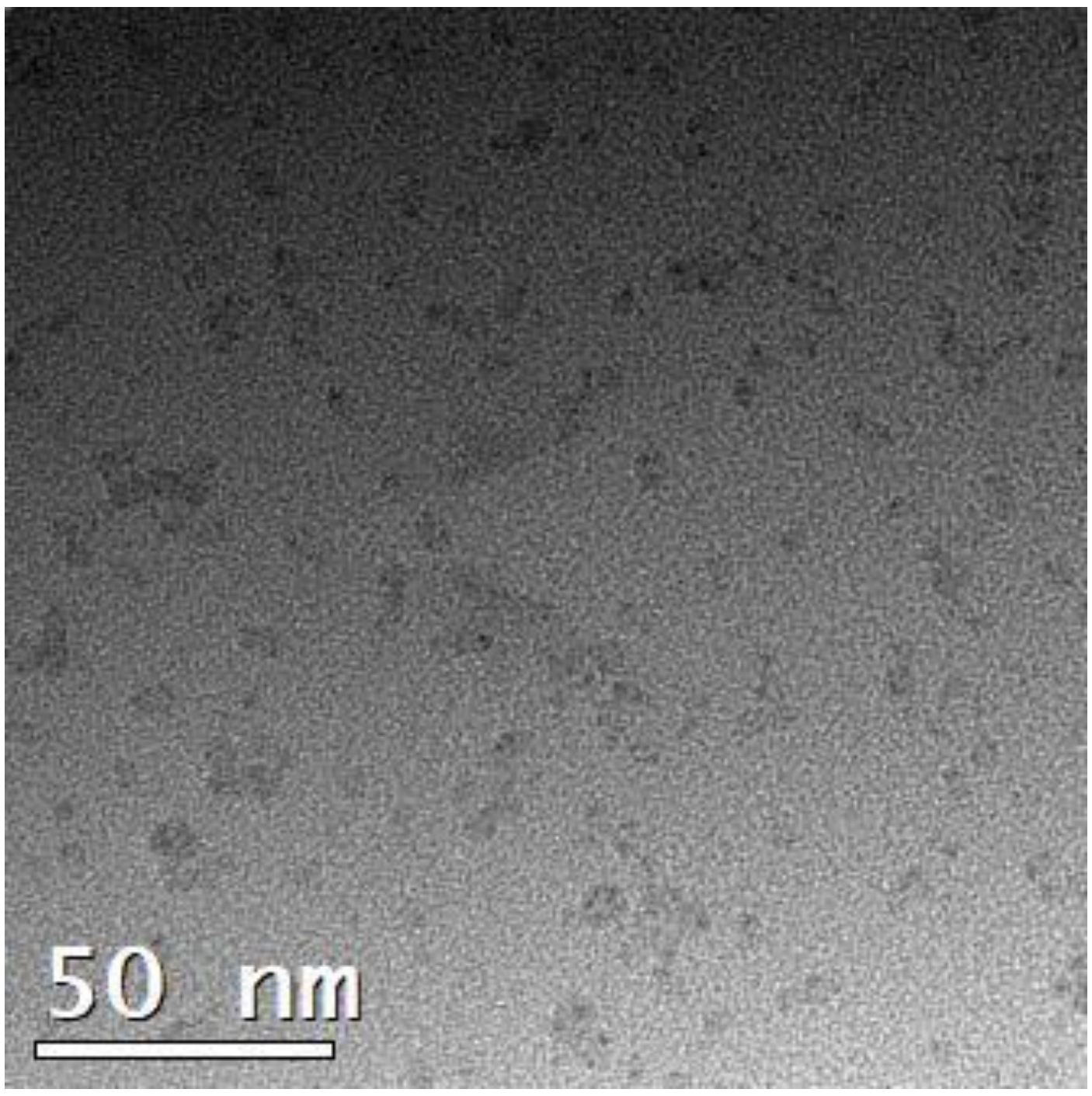
50 nm

C)

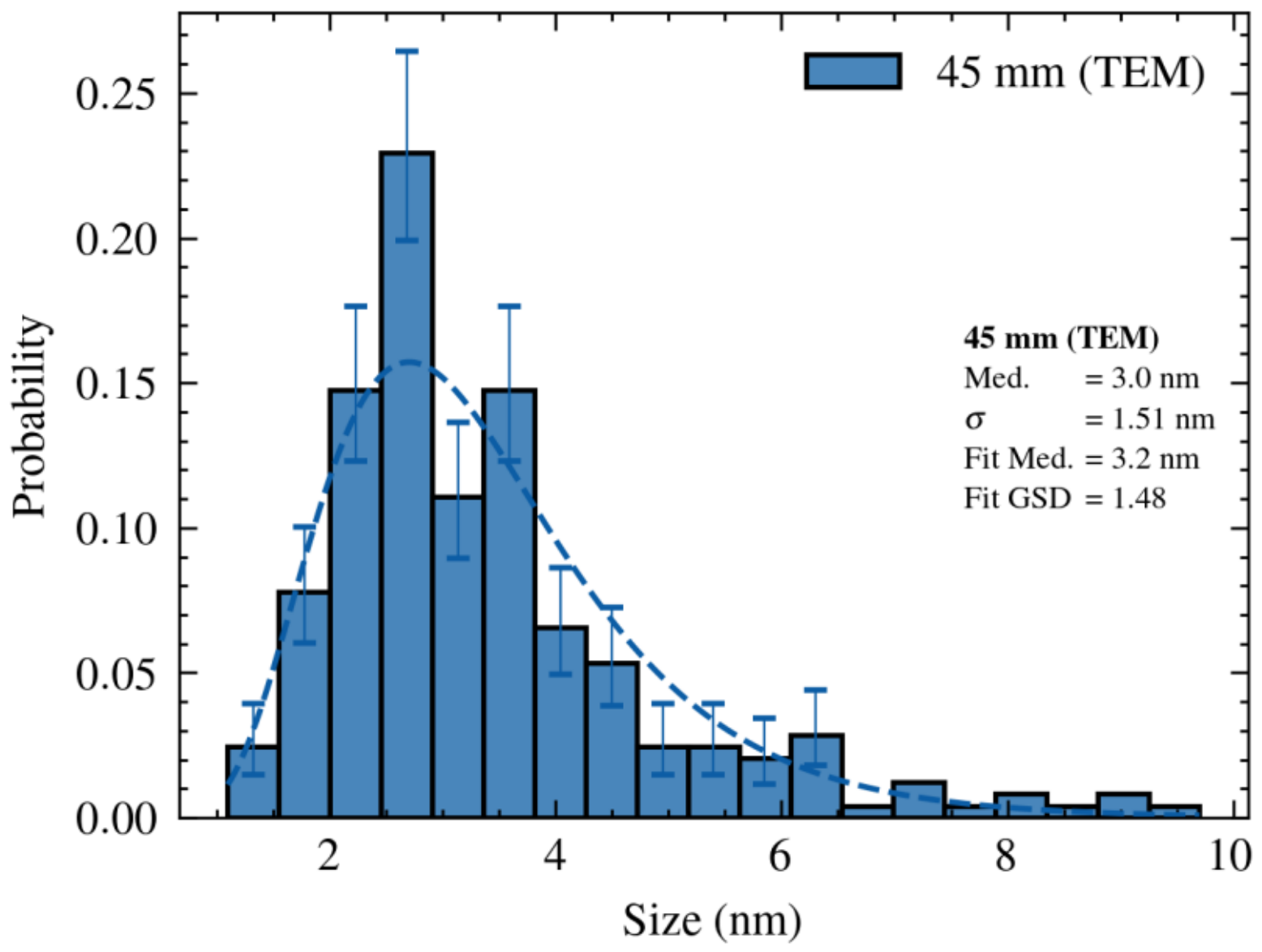

D)

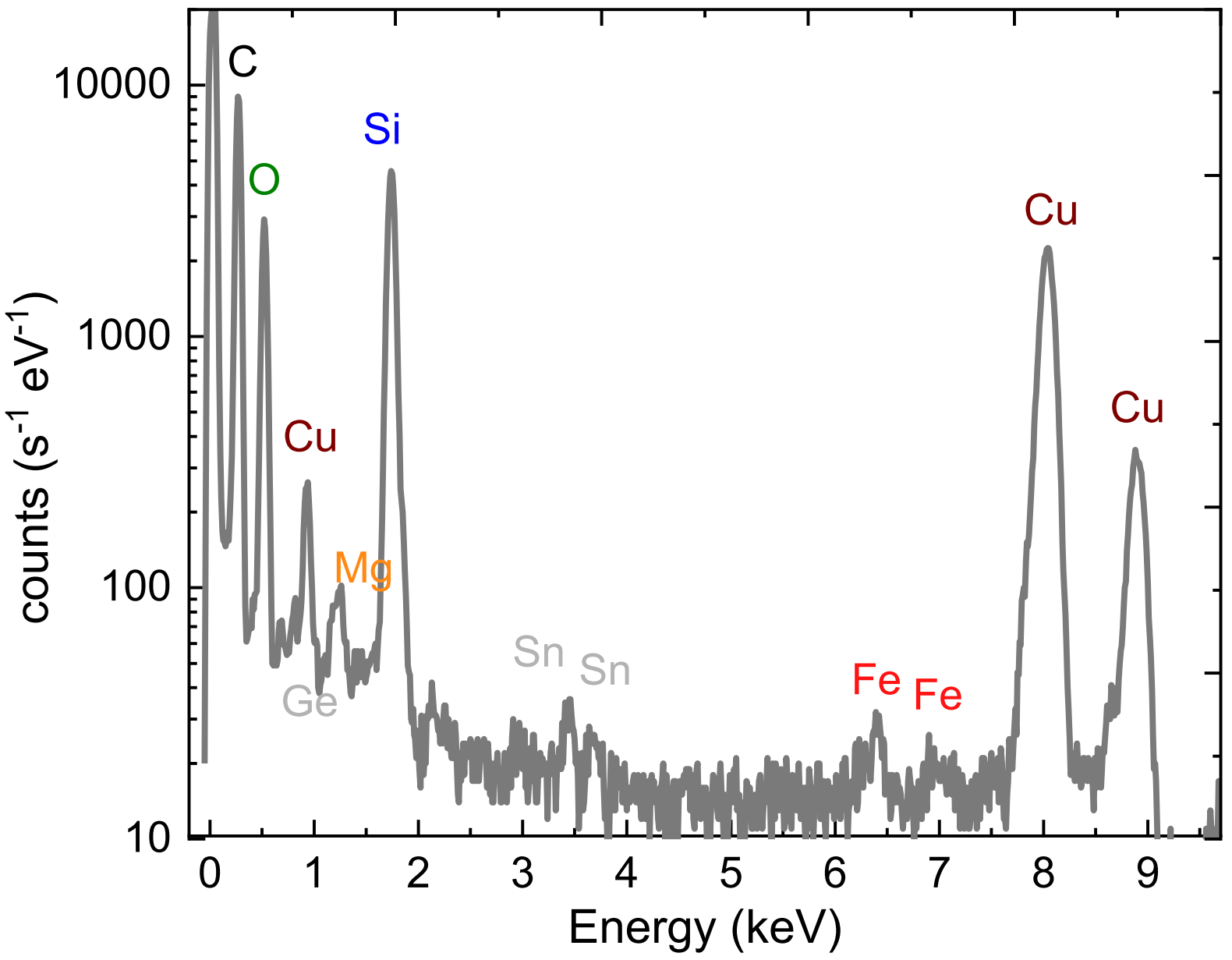

A) schematic depiction of the natural Olivine rock (green disk, 2 inch diameter) inside the magnetron sputtering device to synthesize nanoparticles by the aggregation of emitted Olivine atoms following the Ar ion bombardment in the plasma (pink) B) TEM image of olivine nanoparticles produced from the pristine sputter target made from the natural olivine rock using a 45 mm aggregation distance, 25 sccm gas flow without $H_2$ and 143W magnetron power. C) olivine nanoparticle size distribution from pristine sputter target form B) operated at 45 mm aggregation distance, with average size of 3.0±1.5nm D) EDS of the nanoparticles with Si, O, Fe and Mg confirming olivine properties, Sn and Ge trace contaminations and Cu from the grid.

FIGURE 3 A)

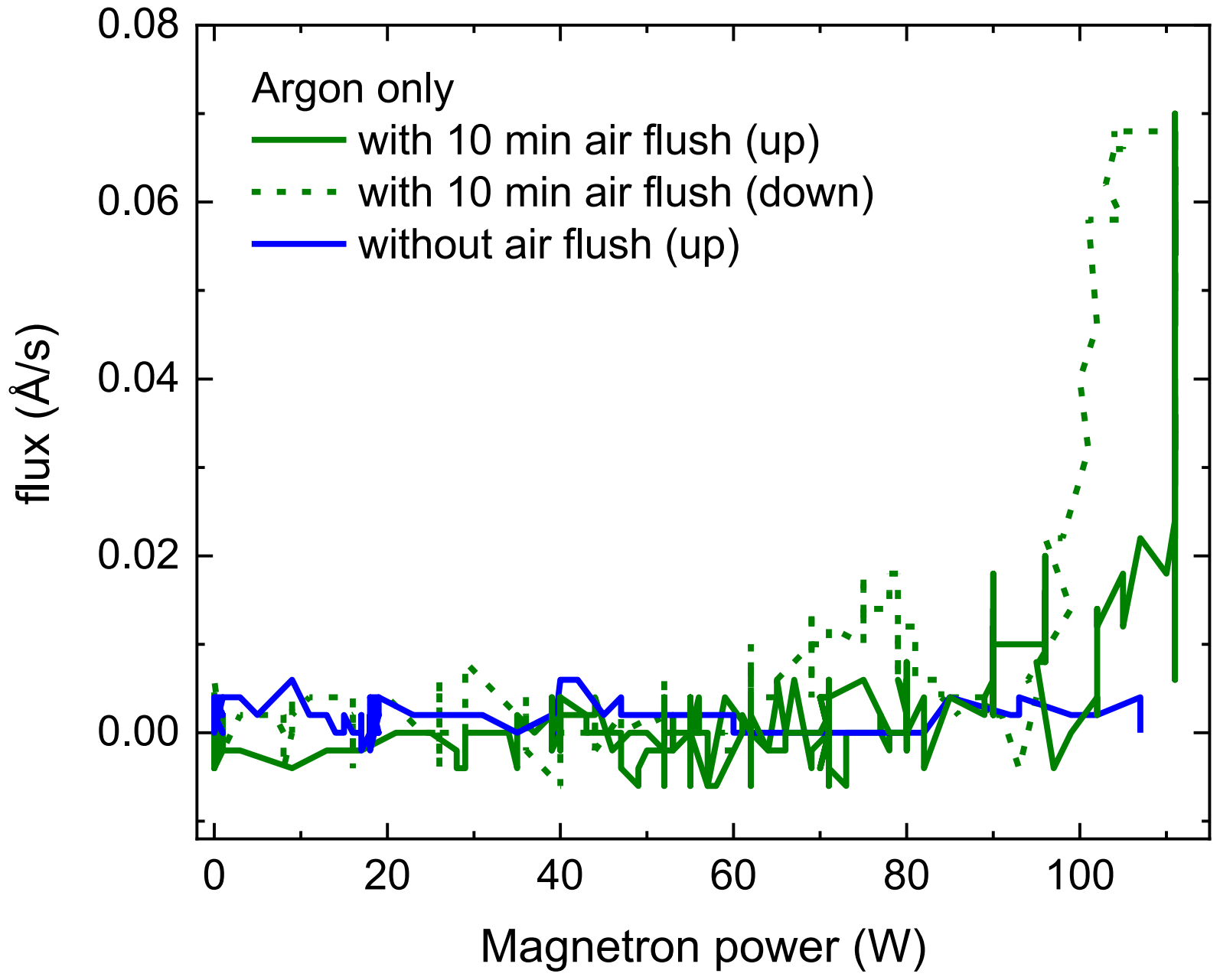
Argon only
with 10 min air flush (up)
with 10 min air flush (down)
without air flush (up)
flux (Å/s)
0.08
0.06
0.04
0.02
0.00
0
20
40
60
80
100
Magnetron power (W)

B)

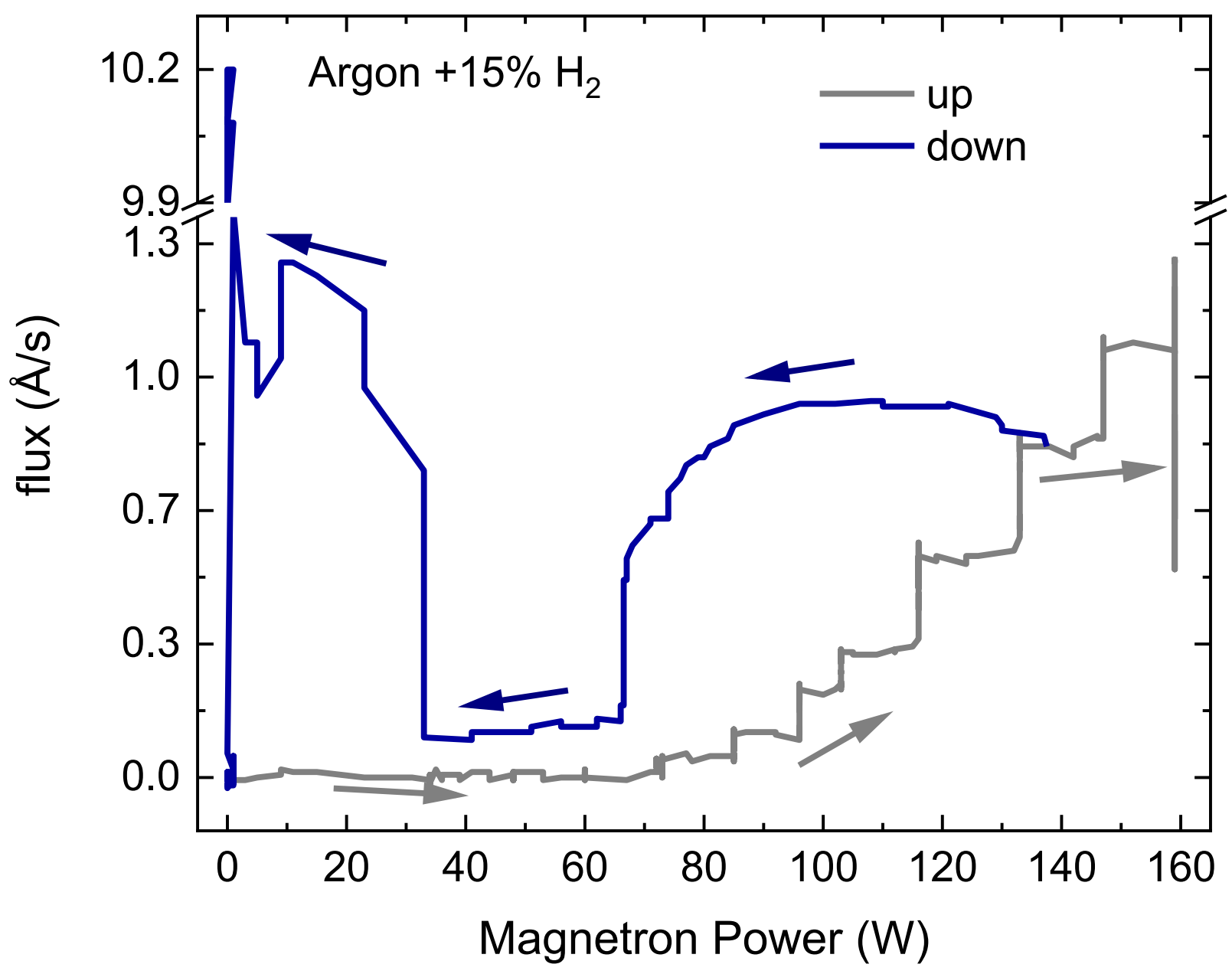
Argon +15% $H_2$
up
down
flux (Å/s)
10.2
9.9
1.3
1.0
0.7
0.3
0.0
0
20
40
60
80
100
120
140
160
Magnetron Power (W)

C)

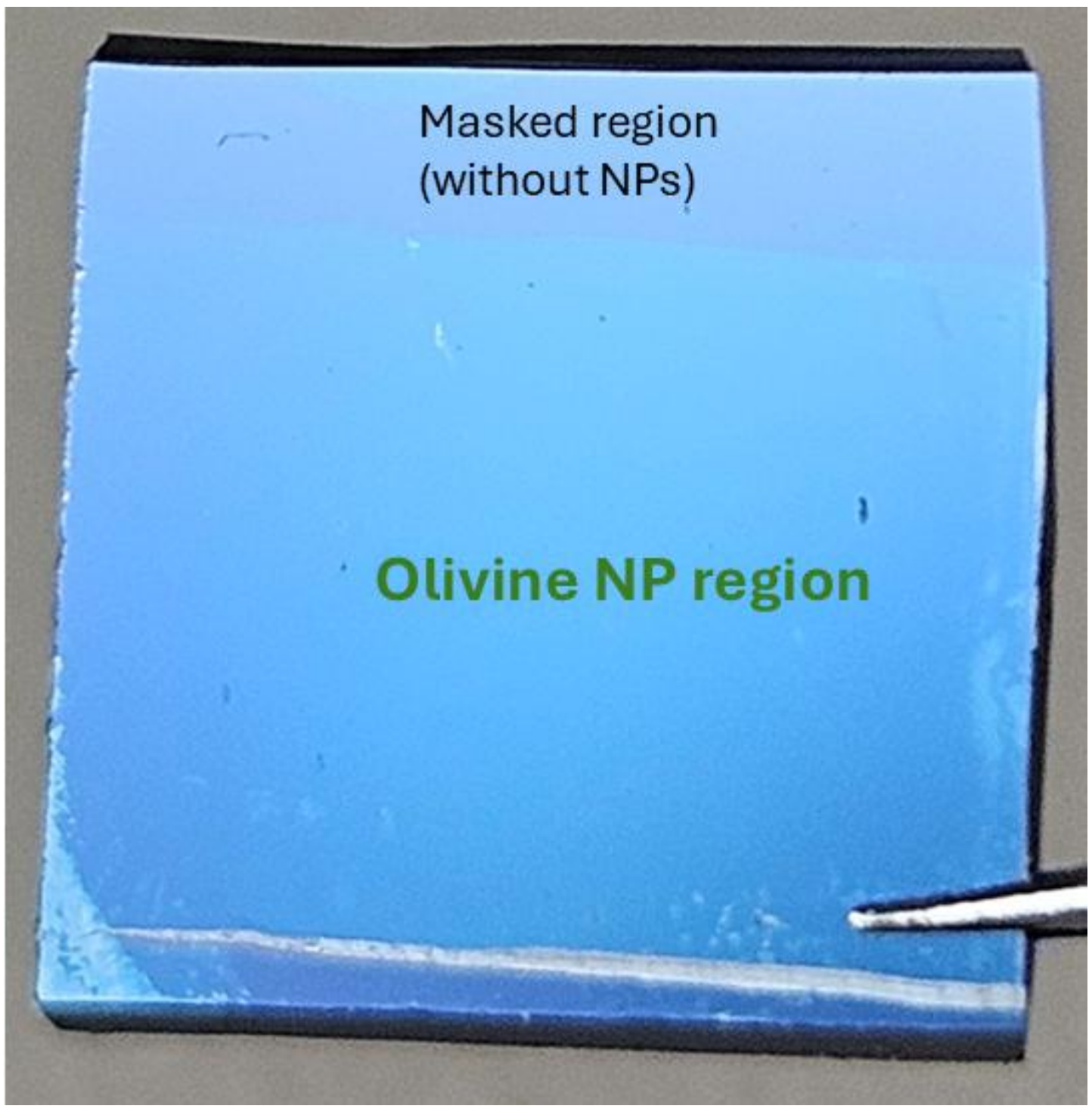

Representative measurements of the Olivine nanoparticle flux (equivalent $SiO_2$ mass) as measured by QCM as function of applied RF magnetron power during ~10 min experiments, in which the power was changed in small steps (~15W ) after stabilisation of the previous step for A) pure argon with negligible nanoparticle flux and with air flush that provided enhanced nanoparticle flux (45 mm aggregation distance and 45 sccm gas flow) and B) with 15% hydrogen (70 mm aggregation distance and 45 sccm gas flow) for going up (grey) to a maximum of 116W and down (blue) in power. A gas flow of 55 sccm was used yielding at a maximum of 116 W a flux of ~1.3Å/s. The flux plateaus are caused by the power step changes. In contrast to the steady flux increase as function of power for the up going curve, the downward curve lingers around a flux of ~0.9Å/s, indicating a sustained plasma that drops at a power around 63W. Around 35W the flux shoots up, suggestive of a new, low energy plasma mode, with a flux peak of ~10Å/s at the few W range. C) olivine nanoparticles deposited on silicon wafer by using air flush and 15% $H_2$. The difference in coloration indicates the deposited and masked regions.

FIGURE 4

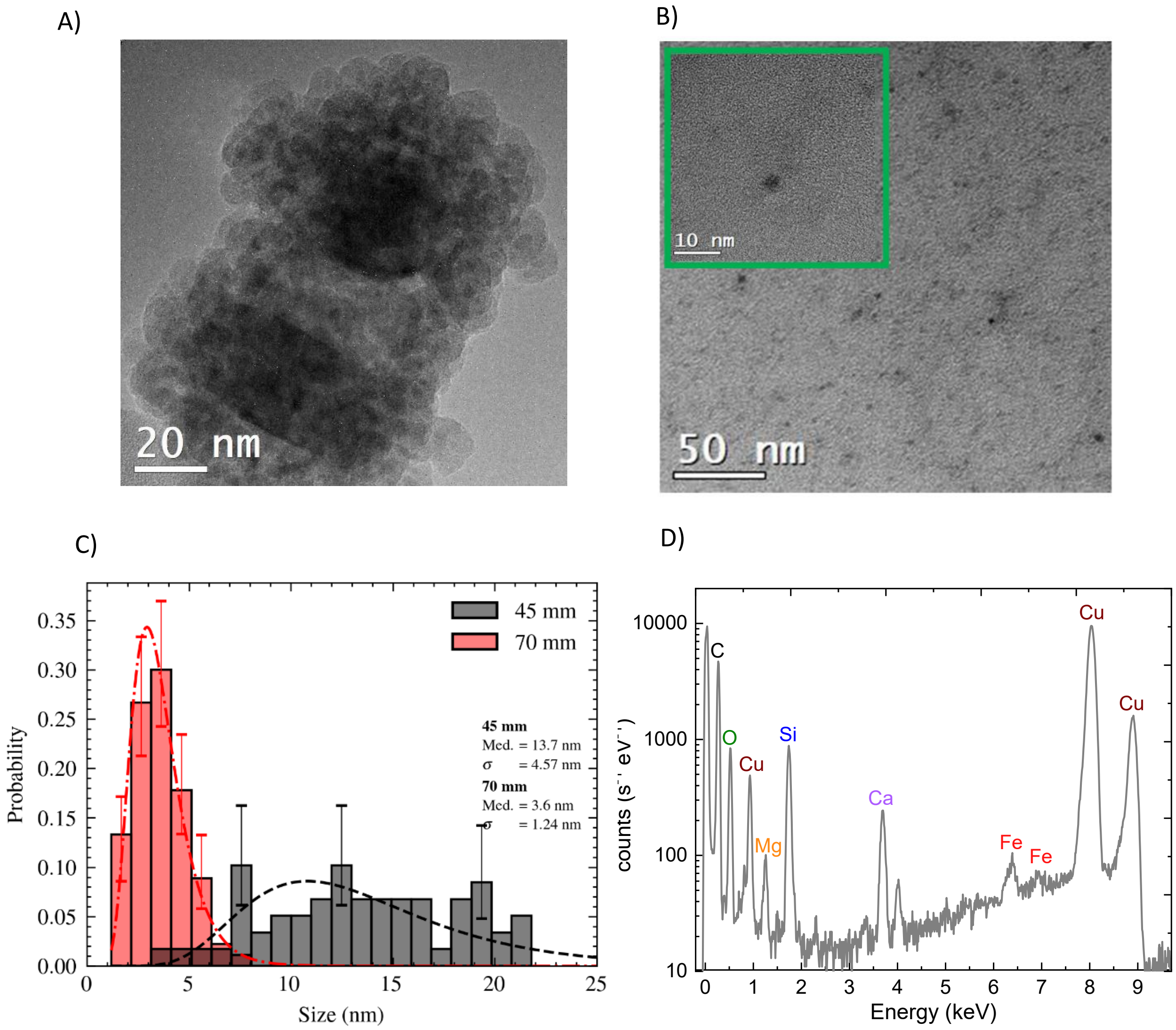

Olivine nanoparticles as imaged by TEM for the a) 45 mm aggregation distance, 45 sccm gas flow (Ar + 15% $H_2$) and 109 W power and b) 70 mm aggregation distances, 30 sccm gas flow flow (Ar + 15% $H_2$) and power of 135 W, with clear differences in terms of particle shape definition and size. The nanoparticles fabricated with the 70 mm aggregation distance appear fainter, likely due to hydroxides, with less contrast. Green box inset of zoom-in. C) size distributions of the olivine nanoparticles with 14±4 nm and 4±1 nm for the 45 mm (grey) and 70 mm (red) aggregation distance, respectively. d) EDX of the nanoparticles (70 mm aggregation length) consistently show the presence of all olivine elements, confirming its complete conversion from the rock, and the presence of substantial amounts of magnesium allows the carbonate formation. Cu originates for the grid support.

FIGURE. 5

A)

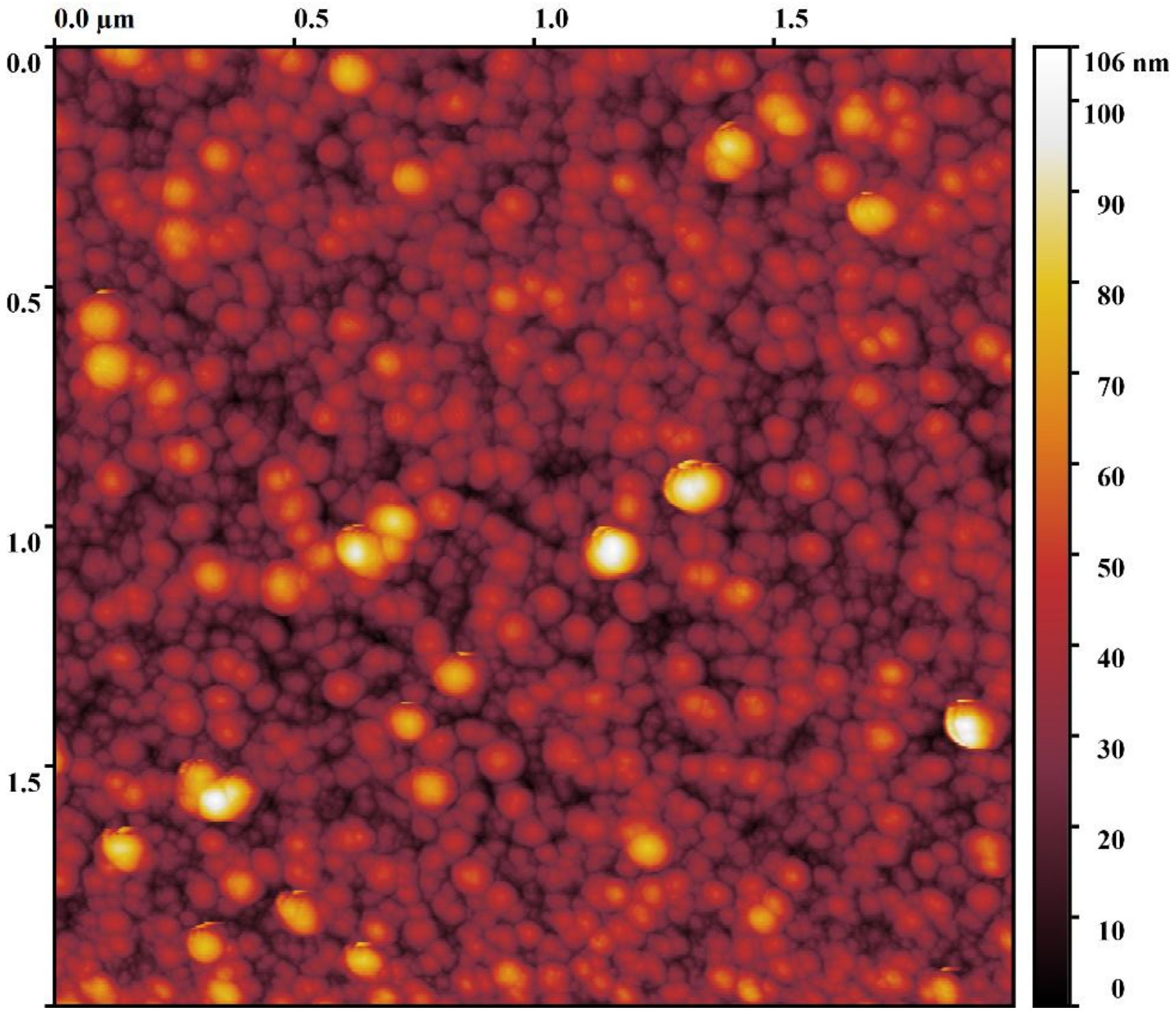

B)

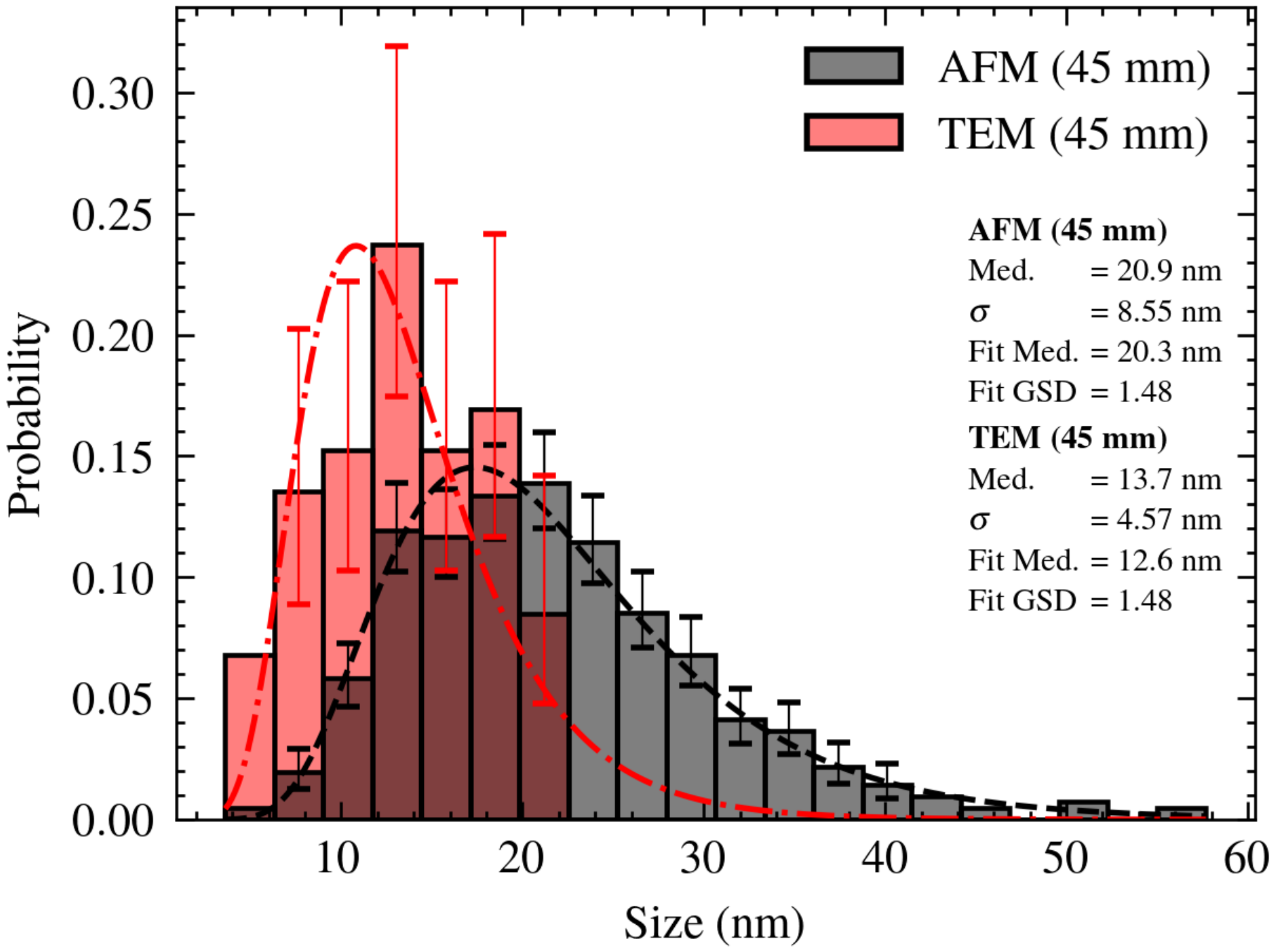

A) Natural olivine nanoparticles deposited on a silicon wafer with aggregation length of 45 mm, 75 sccm gas flow (Ar + 15% $H_2$) and power of 106 W, as imaged by AFM, with distinctive single nanoparticles, forming a nanoparticle assembled layer. From the height of the nanoparticles the size could be determined. B) size distribution of with Log Normal fitting providing mean diameters 21±9 nm and 10 ±4 nm for the aggregation length 45 mm(grey) and 70 mm (red), respectively as obtained from various samples at locations with low coverage.

TABLE 2.

| Element | Concentration (at. %) | |
|---|---|---|
| | **45 mm** | **70 mm** |
| O | 42 | 39 |
| Mg | 7 | 10 |
| Fe | 1.5 | 1.5 |
| Si | 7 | 4.2 |
| Ge | 1.5 | 1.2 |
| C | 40 | 40 |
| Trace elements* | 0.4 | 0.7 |
| Other contaminants** | 0 | 2.2 |

Composition of as received samples deposited with aggregation distance 45 and 70 mm, retrieved by X-ray Photoelectron Spectroscopy (XPS). *refer to inclusions of chromium and tin, ** refer to traces of sulfur and tungsten.

FIGURE 6

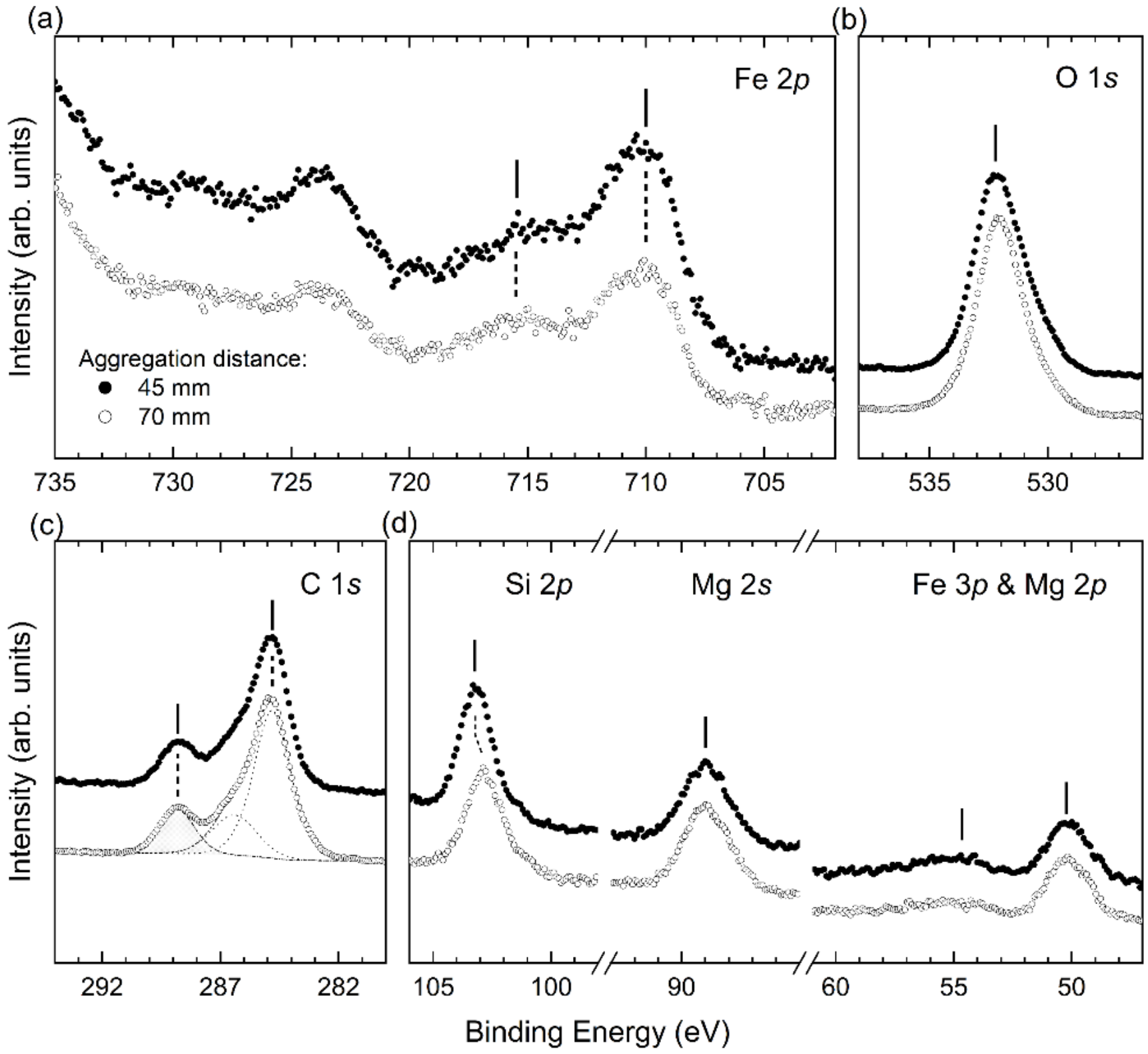

X-ray photoelectron spectra (Mg *K*α photons) acquired from as-received samples with aggregation length 45 and 70 mm. A numerical fit is performed on the spectrum from the latter sample, using Voigt-like line shapes. Source satellites have been removed from the experimental data. Vertical lines highlight the Binding Energy (BE) position of specific features described in the main text. Panel (d) shows excerpts from the same wide-range spectra, allowing a direct comparison of the relative feature intensities.

FIGURE 7

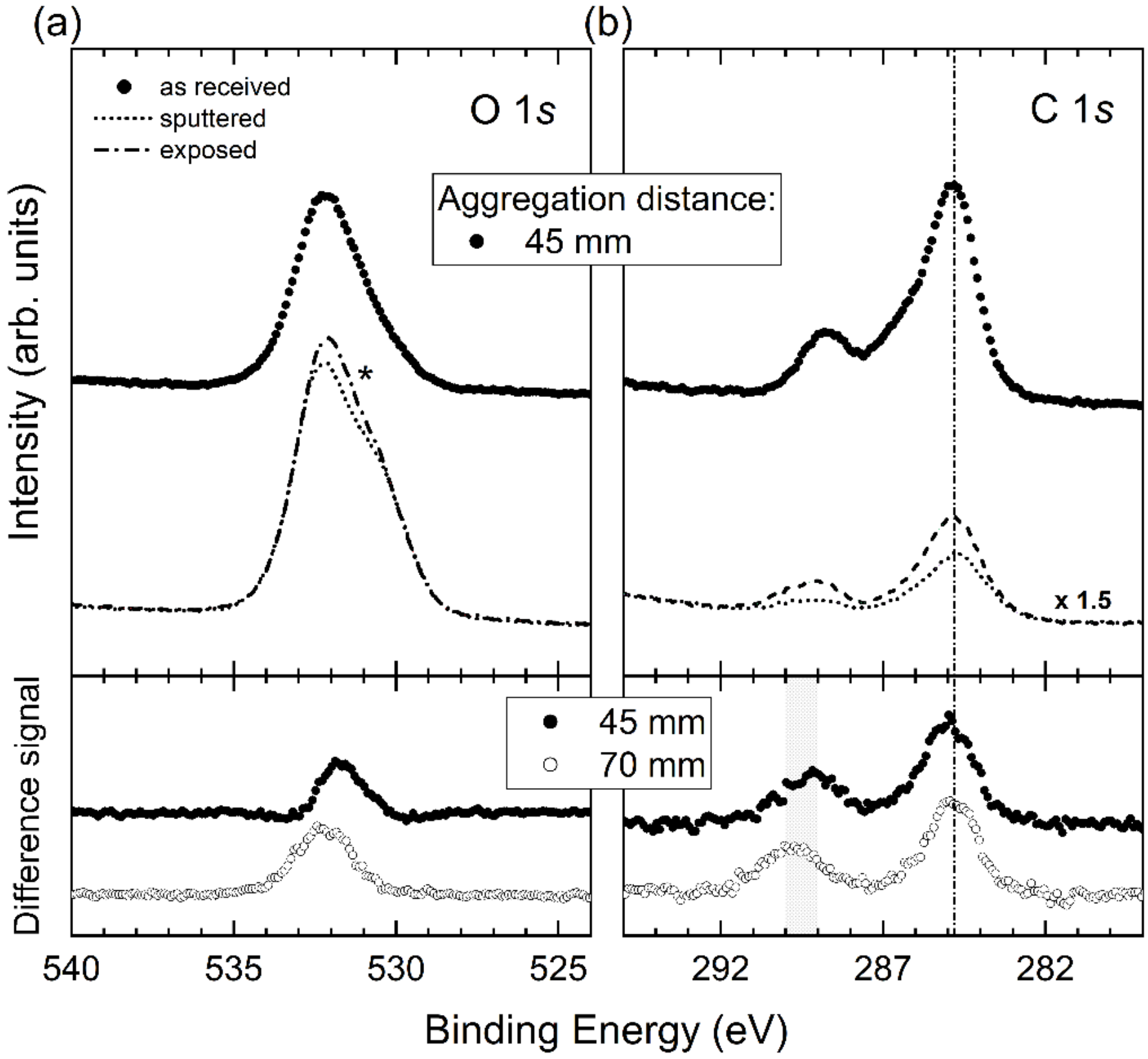

Photoemission line shape evolution upon sputtering and $CO_2$ exposure, considering a) the O 1*s* and b) the C 1*s* BE regions. The asterisk marks the position of a shoulder of the main O 1*s* feature that develops upon sputtering, the vertical line marks the reference BE of hydrocarbon-like species in the C 1*s* BE region. Top panels: spectra acquired from a sample produced by choosing an aggregation length of 45 mm; bottom panels: difference spectra (exposed − sputtered) for samples with aggregation lengths of 45 and 70 mm, after normalization to the Mg signal intensity. The carbonate signal falls within the shaded area in panel (b), bottom spectra.

FIGURE 8

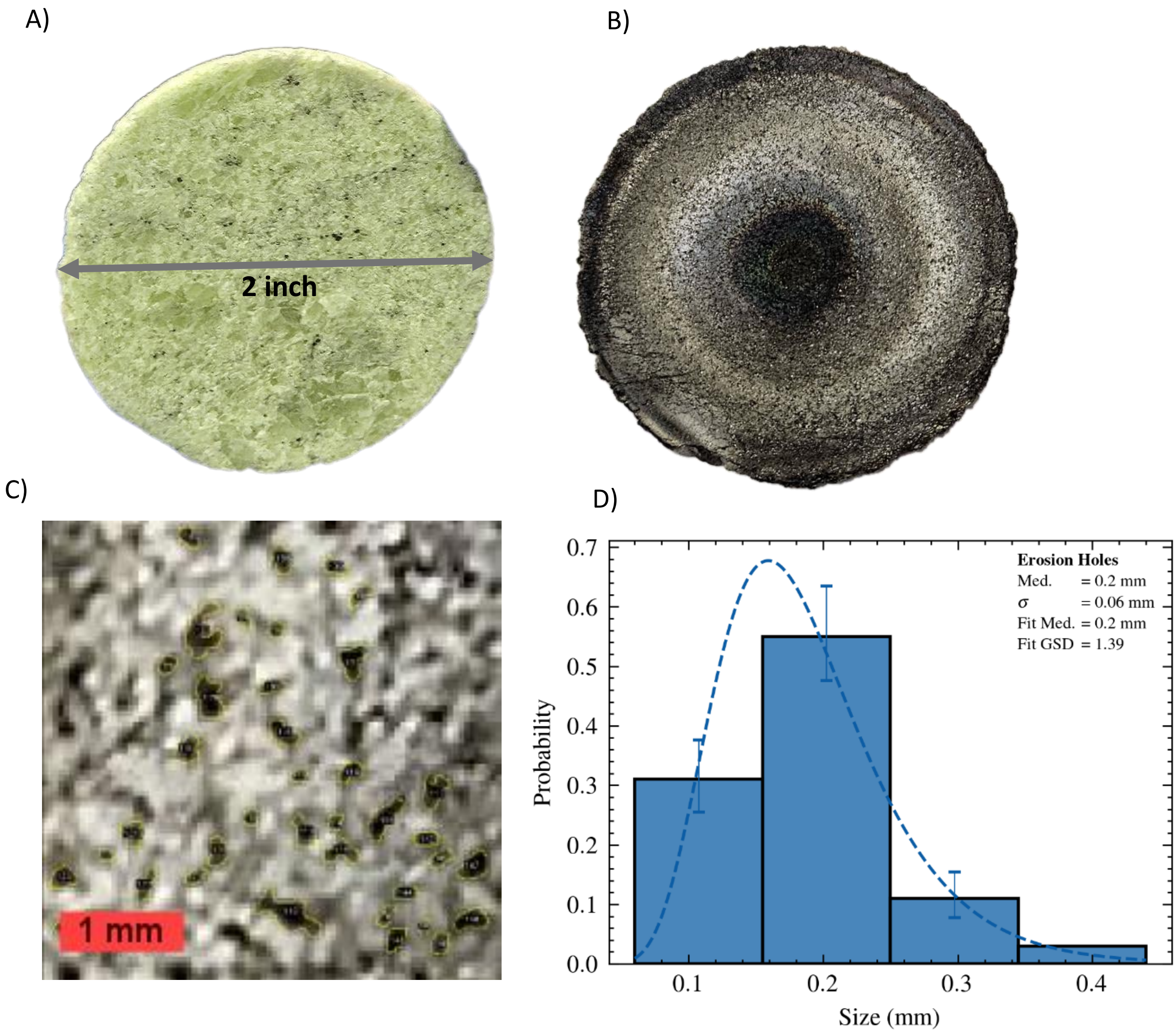

A) Olivine sputter target A) before and B) after sputtering, with clear macroscopic voids in the surface due to the sputter erosion of softer parts of the Olivine. The voids are shown in C) and D) their size distribution as the area is approximated by a circle, is provided as a histogram that could be well fitted with a Log norm function with a mean diameter of ~0.2±0.1 mm. The voids are created by enhanced plasma erosion due to fragments with lower binding energy.

FIGURE 9

A)

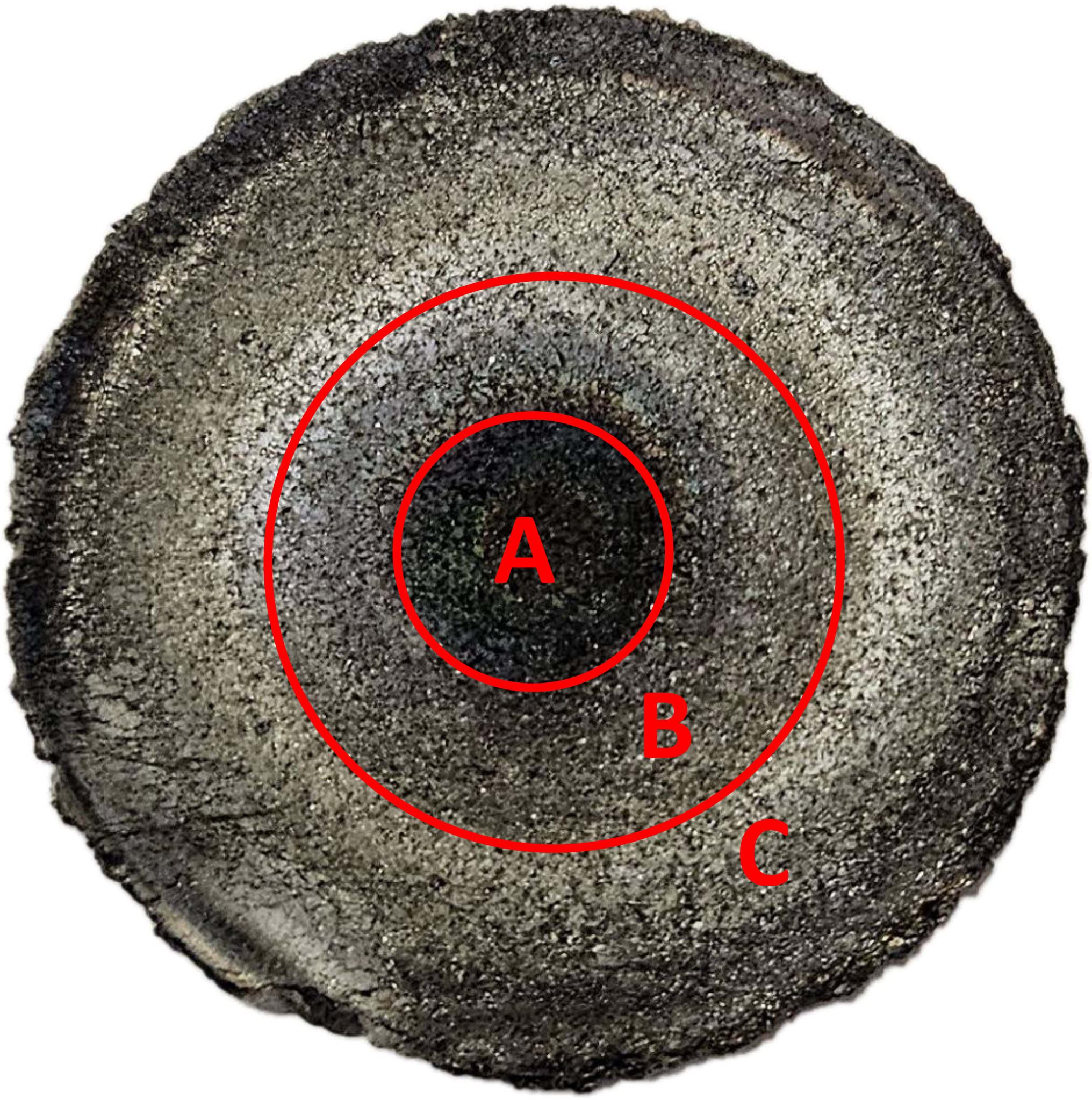

B)

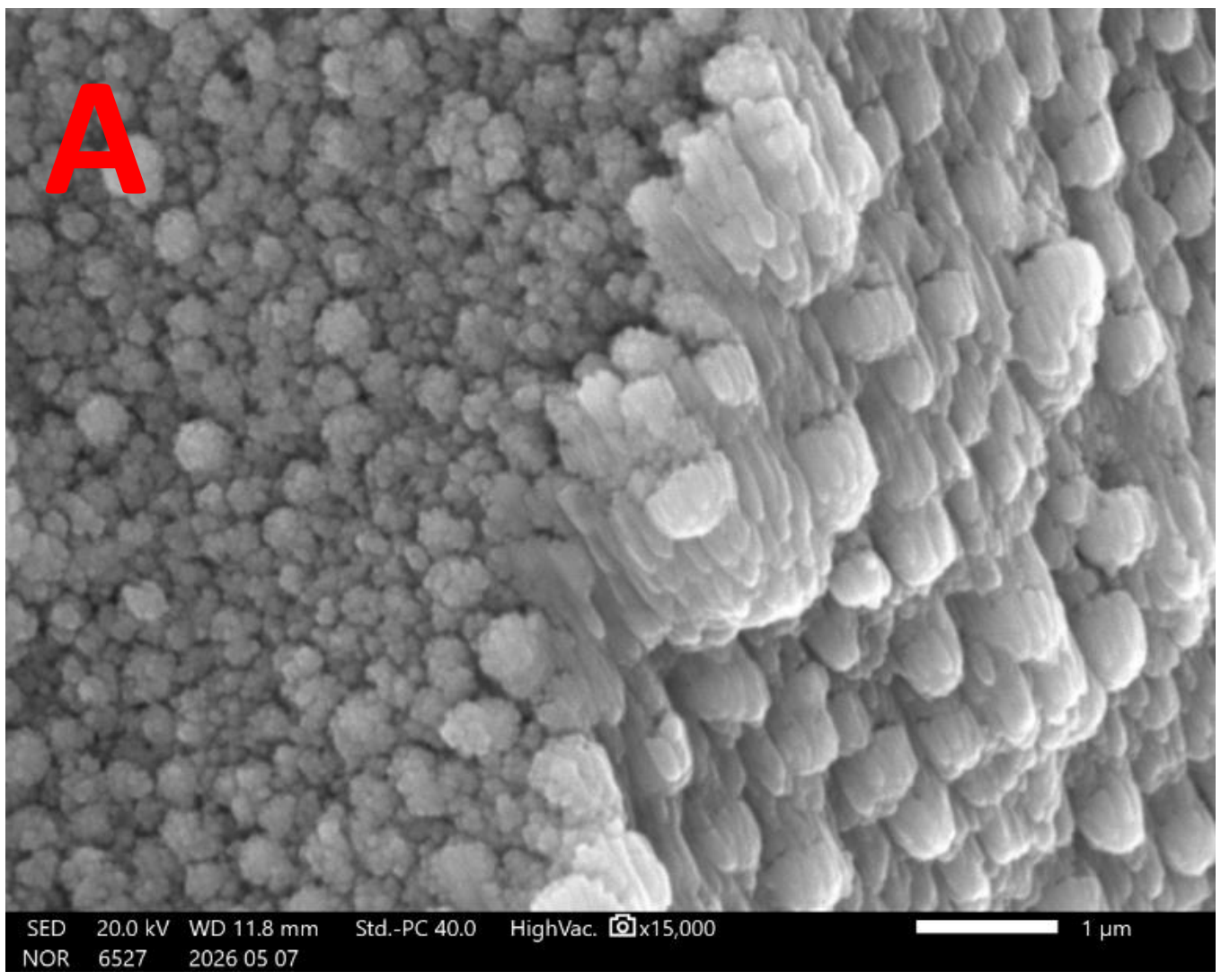

C)

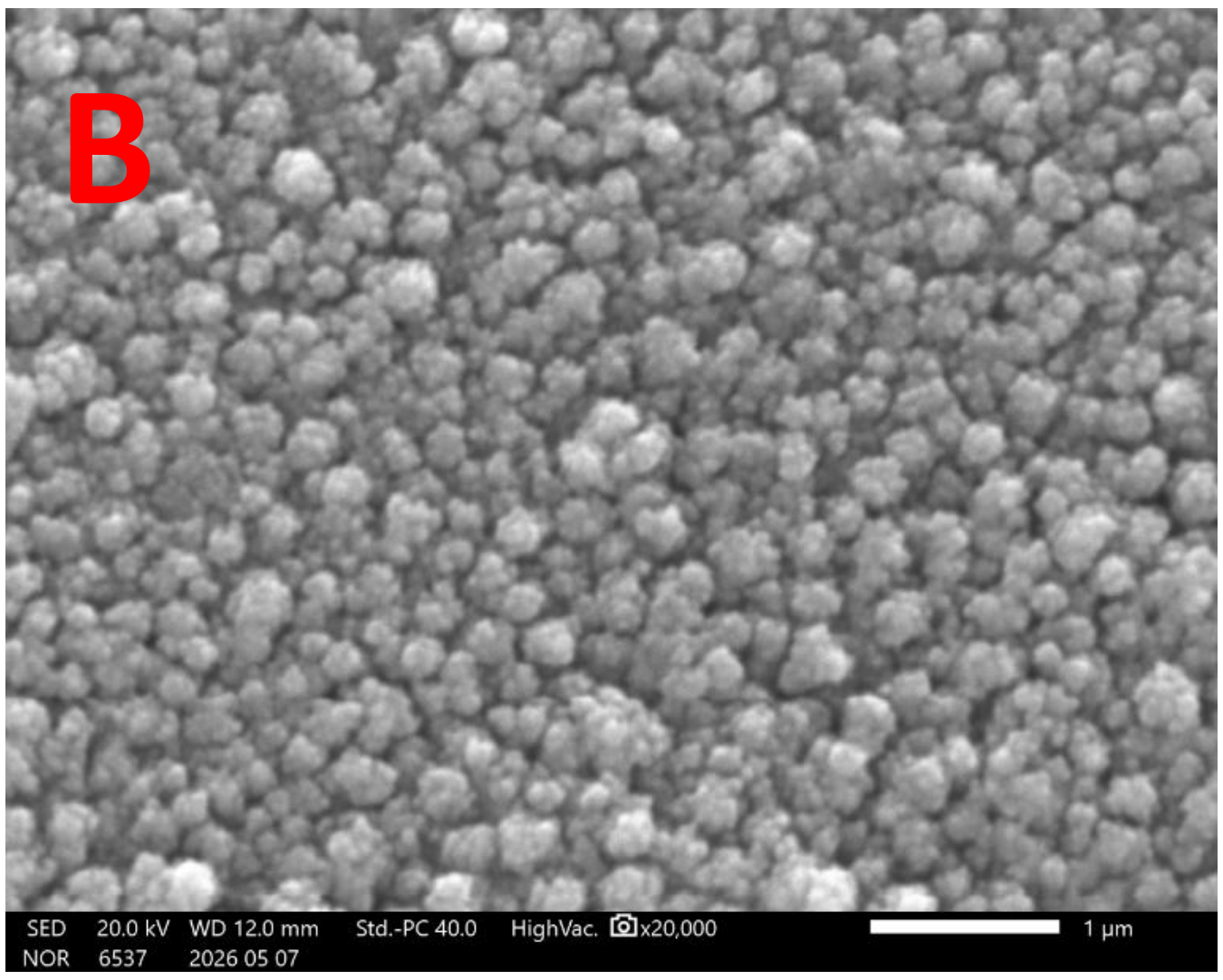

A) Eroded olivine sputter target with three distinct areas indicated by different colouration. B) nanostructures with ~100 nm features from spot A and C) similar nanostructures from spot B.

FIGURE 10

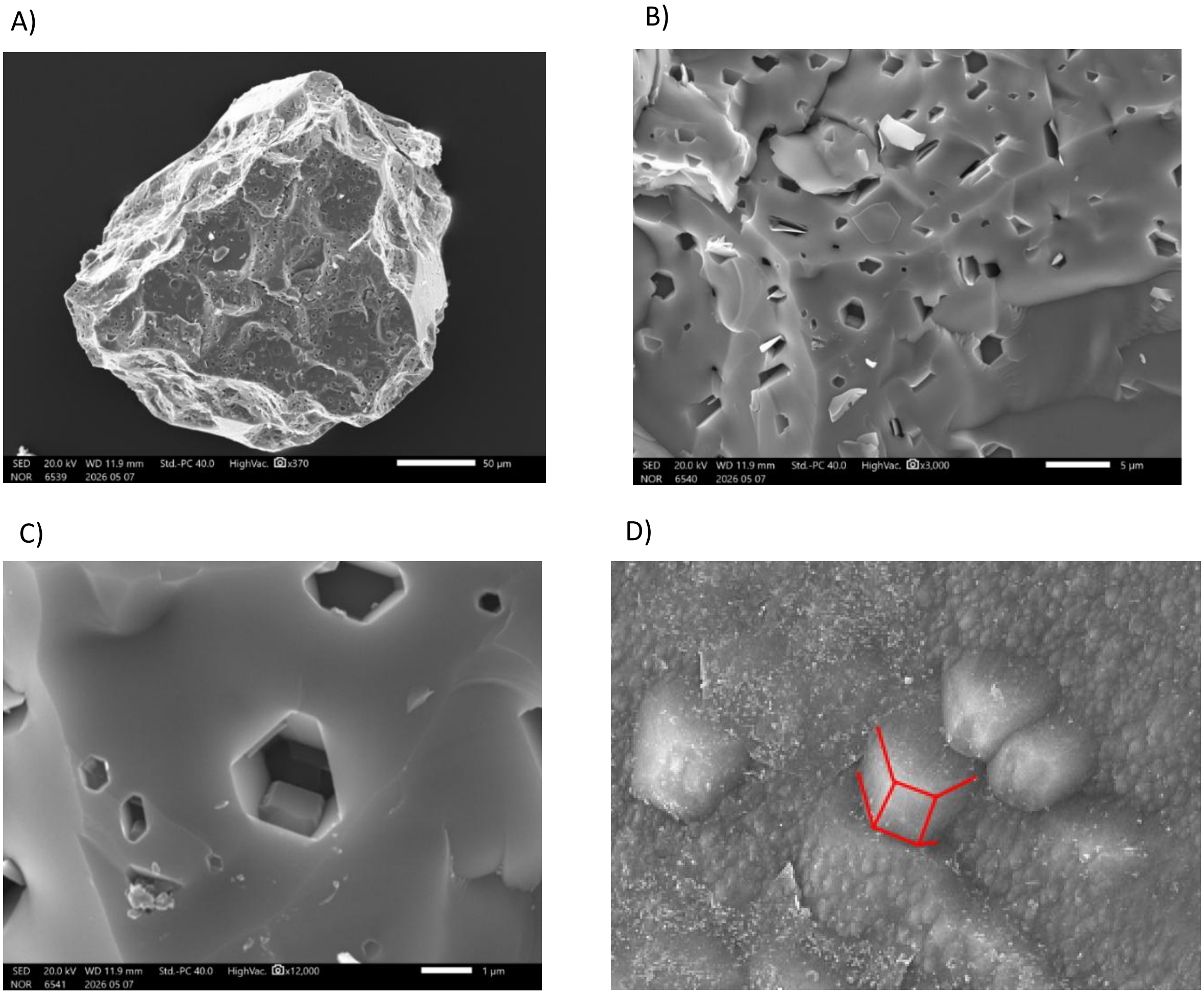

Olivine sputter target fragment from A) position B that has been etched by the Argon plasma, leaving distinct, geometrically shaped voids at positions which likely contained crystals with higher sputter yield due to lower chemical bonding. In B) and C) zoom in displaying the remarkable well defined geometrical holes. D) Protruding truncated pyramid crystals from the sputter target position C with likely lower sputter yield due to the highly crystalline nature and hence stronger chemical bonding than the surrounding Olivine. EDS analysis on the truncated pyramids provided atomic concentrations for Mg: 24%, Fe: 1.7%, Si: 13% and O: 55%, which were very similar off the truncated pyramids.

FIGURE 11

A)

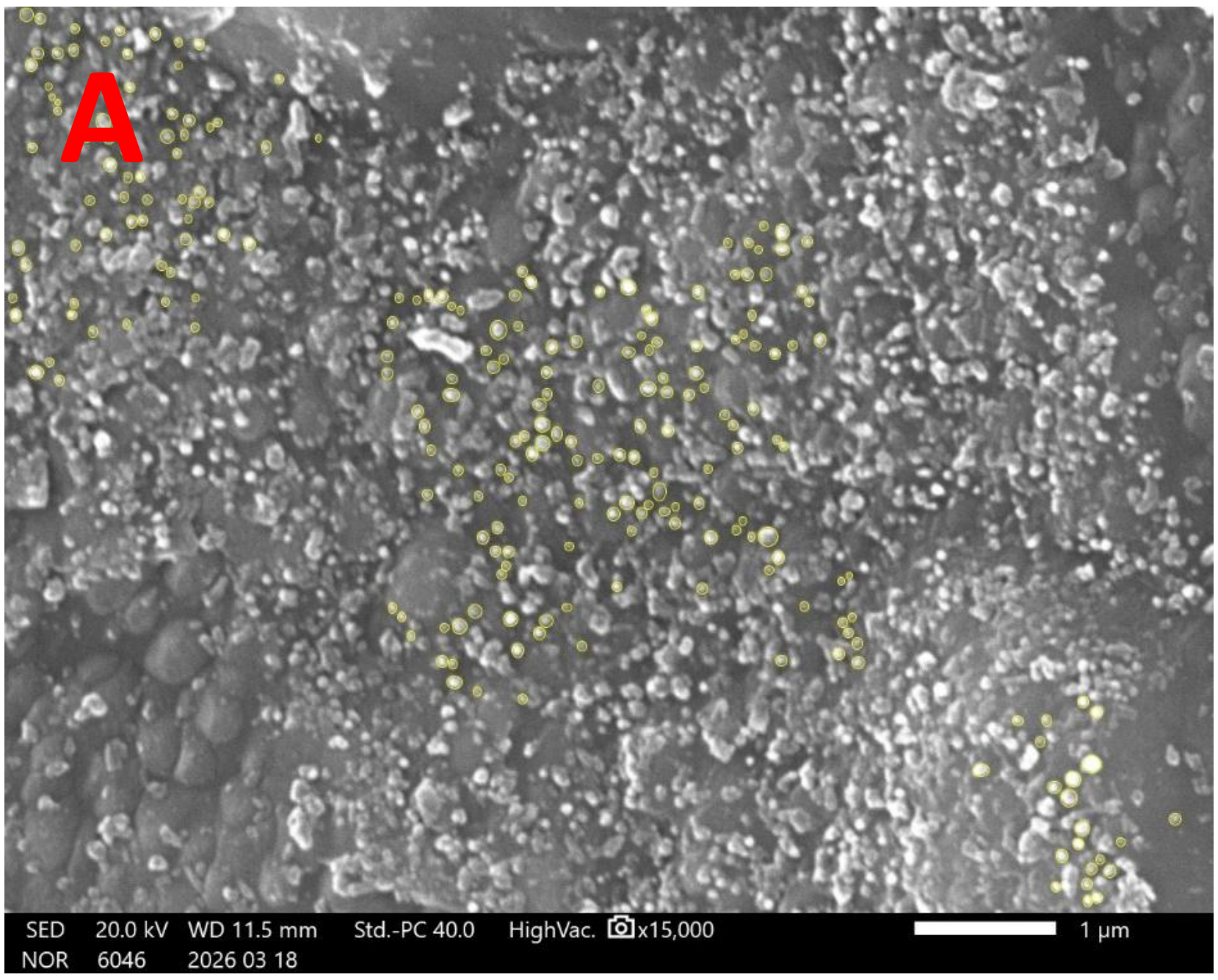

B)

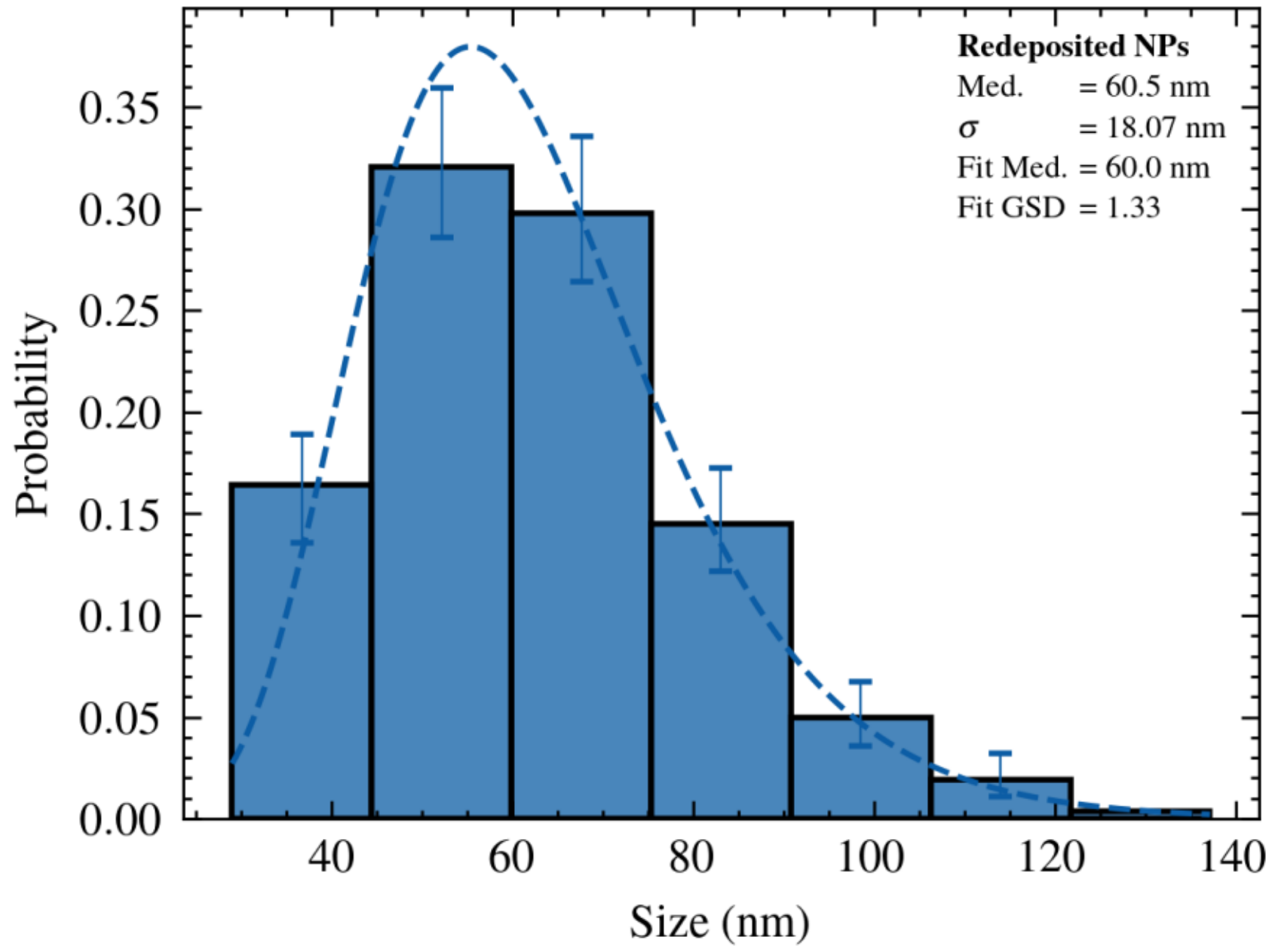

A) Redeposited olivine nanoparticles (yellow spots indicated) on central region of sputter target B) redeposited nanoparticle size distribution with mean particle size of 72±1 nm, which is at the upper size range as measured by AFM. EDS elemental analysis provided atomic concentration for Mg: 1.7%, Fe: 34%, Si: 1.4% and O:10%, the clear presence of Olivine constituents with a higher concentration of iron, perhaps due to its interaction with the magnetic field. The presence of germanium (5%) on the sputter target originates from the previous deposits on the reactor chamber wall, while Chromium likely was present in the olivine rock.